\documentclass[3p,number,sort&compress]{elsarticle}
\usepackage{color}
\usepackage{graphicx}
\usepackage{amsmath,amssymb}
\usepackage{slashed}

\DeclareMathOperator{\sign}{sign}
\DeclareMathOperator{\Pf}{Pf}

\newcommand{\ZZ}{\mathcal{Z}}
\newcommand{\Dd}{\mathcal{D}}
\newcommand{\abs}[1]{\left| #1 \right|}
\newcommand{\SB}{S_\text{B}}
\newcommand{\vev}[1]{\left\langle #1 \right\rangle}
\newcommand{\bbZ}{\mathbb{Z}}
\newcommand{\trnsp}{\mathsf{T}}
\newcommand{\Nt}{N_\text{t}}
\newcommand{\Ns}{N_\text{s}}
\newcommand{\ord}{\mathcal{O}}
\newcommand{\Seff}{S_{\text{eff}}}
\newcommand{\half}{{\textstyle\frac{1}{2}}}
\newcommand{\ee}{\mathrm{e}}
\newcommand{\dd}{\mathrm{d}}
\newcommand{\DD}{\mathrm{D}}
\newcommand{\const}{\mathrm{const.}}
\newcommand{\ket}[1]{\vert #1 \rangle}
\newcommand{\bra}[1]{\langle #1 \vert}
\newcommand{\braket}[2]{\langle #1 \vert #2 \rangle}
\newcommand{\sch}{\mathcal{Q}}

\DeclareMathOperator{\Tr}{Tr}

\graphicspath{{plots/}}

\journal{arXiv}
\submitDate{July 17, 2011}
\begin{document}
\title{Supersymmetry Breaking in Low Dimensional Models}

\author{Christian Wozar}
\ead{christian.wozar@uni-jena.de}
\author{Andreas Wipf}
\ead{wipf@tpi.uni-jena.de}
\address{Theoretisch-Physikalisches Institut,
Universit{\"a}t Jena, D-07743 Jena, Germany}

\begin{abstract}
\noindent
We analyse supersymmetric models that show supersymmetry breaking in
one and two dimensions using lattice methods. Starting from supersymmetric
quantum mechanics we explain the fundamental
principles and problems that arise in putting supersymmetric models 
onto the lattice. We compare our lattice results (built upon 
the non-local SLAC derivative) with numerically exact
results obtained within the Hamiltonian approach. A particular
emphasis is put on the discussion of boundary conditions. We investigate
the ground state structure, mass spectrum, effective potential
and Ward identities and conclude that lattice methods are suitable
to derive the physical properties of supersymmetric quantum mechanics, 
even with broken supersymmetry.
Based on this result we analyse the two dimensional $\mathcal{N}=1$
Wess-Zumino model with spontaneous supersymmetry breaking. First we 
show that (in agreement with earlier analytical and numerical studies) 
the SLAC derivative is 
a sensible choice in the quenched model, which is nothing but the two 
dimensional $\phi^4$ model. Then, we present the very first computation of a
renormalised critical coupling for the complete supersymmetric model. 
This calculation makes use of
Binder cumulants and is supported by a direct comparison to Ward identity
results, both in the continuum and infinite volume limit. The physical picture
is completed by masses at two selected couplings, one in the supersymmetric phase and one in the supersymmetry broken
phase. Signatures of the Goldstino in the fermionic correlator are clearly
visible in the broken case.
\end{abstract}

\maketitle

\section{Introduction}
\label{chap:intro}

\noindent
Symmetries are one of the guiding principles in contemporary theoretical physics.
They led to the construction of the standard model of
electroweak interaction \cite{Glashow:1961tr,Salam:1964ry,Weinberg:1967tq} 
and strong interaction \cite{Fritzsch:1973pi}. Based on the underlying 
symmetries bottom and top
quark as well as the $\tau$ neutrino have been predicted and the experimental
discoveries, the last one more than two decades after the prediction
\cite{Kodama:2000mp}, substantiated the success of the standard model to
describe the physics on energy scales below $1$~TeV.
The extension of the standard model's symmetries by further ones is 
constrained by the celebrated Coleman-Mandula theorem
\cite{Coleman:1967ad}. A way to circumvent this no-go theorem 
is given by extending the Poincar\'e algebra with anti-commuting
\emph{supersymmetry} generators \cite{Haag:1974qh} that relate particles with
integer spin to ones with half-integer spin.

The first field theoretical realisation of a renormalisable model with
supersymmetry is the Wess-Zumino model \cite{Wess:1973kz}
in four spacetime dimensions with a field content of two (real) scalars and a
Majorana fermion. Since then a variety of models with global supersymmetry 
have been constructed, for example supersymmetric gauge theories or supersymmetric 
sigma models and some of these possess several supersymmetries. 
For an introduction, see \cite{Sohnius:1985qm,Martin:1997ns,Wess:1992cp}. 
Supersymmetric extensions of the well established standard model 
may solve or weaken several problems of the standard model, such as the 
hierarchy problem, the occurrence of dark matter and the strong $\mathrm{CP}$ 
problem \cite{Dimopoulos:1981zb,Ellis:1983ew,Hiller:2001qg}.

Supersymmetric models have certain theoretical advantages, such as less severe
divergences in perturbation theory as compared to models without supersymmetry, and
the supersymmetry algebra induces a vanishing ground state energy, as long
as supersymmetry is unbroken. If the ground state is invariant under
supersymmetry, mass degenerate multiplets of bosonic and fermionic particles are
predicted, and it known that in certain classes
of supersymmetric theories a spontaneous breaking of supersymmetry is not
possible \cite{Witten:1982df}. 
In experiments no such mass degeneracy has been detected so far, and the masses 
of bosonic and fermionic particles appear to be unrelated. At first sight
these results tell us that supersymmetry is not realised in nature.
However, as first analysed by O'Raifeartaigh \cite{ORaifeartaigh:1975pr} this
non-degeneracy of masses is naturally expected if supersymmetry is
\emph{dynamically broken}.

In a theory with dynamical supersymmetry breaking the ground state is not
invariant under supersymmetry, and the ground state energy is lifted above
zero \cite{Witten:1981nf}. But the supersymmetry algebra is still present
and this has implications for the physics in the dynamically broken sector.
To date the Large Hadron Collider is fully operating 
and it is expected to measure remnants of supersymmetry in collision events
within the upcoming years. Clearly, if supersymmetry plays any role
in nature then it is mandatory to explore supersymmetric theories with 
methods that are applicable in the non-perturbative regime.

Among those methods the lattice regularisation in combination
with importance sampling based statistical `Monte-Carlo' methods has been 
most successful over the last decades. Lattice methods often provide the only
viable way to gain information about the non-perturbative sector of quantum
field theories. Early simulations that aimed at an understanding of
the pure $SU(2)$ Yang-Mills theory \cite{Creutz:1980zw} built the basis for
recent computations from first principles of the Hadron spectrum in full
quantum chromodynamics \cite{Durr:2008zz}, which is only possible due to
increasing computing power and algorithmic improvements. As non-perturbative effects
are automatically taken into account in lattice simulations, it is desirable 
to apply the lattice approach to supersymmetric theories as well. This has 
been the subject of a number of publications, see, e.g.,
\cite{Feo:2004kx,Giedt:2006pd,Takimi:2007nn,Damgaard:2007eh,Catterall:2007kn}
and for recent progress in supersymmetric Yang-Mills theories
\cite{Elliott:2008jp,Giedt:2009yd,Endres:2009yp,Demmouche:2010sf} and references
therein. 

In all lattice regularised versions of field theories symmetries are
of particular interest. If a symmetry of the continuum theory is not
implemented in the lattice version it may happen that the symmetry is not restored in 
the continuum limit. E.g.\
for simulations of gauge theories it is important to implement the
lattice version of the continuum gauge symmetry \cite{Wilson:1975id}. But
not every symmetry can be directly implemented in the lattice regularised
theory. For instance, the Nielsen-Ninomiya theorem
\cite{Nielsen:1980rz,Nielsen:1981xu,Nielsen:1981hk} forbids the exact implementation of 
chirally symmetric fermions with a local fermion interaction
and without introducing additional fermion flavours on the lattice.
Nevertheless, it is possible to construct a (deformed) lattice version of the
chiral symmetry, which is given by the Ginsparg-Wilson relation
\cite{Ginsparg:1981bj}, so that a restoration of the continuum chiral symmetry
is ensured in the continuum limit of the lattice action.

For supersymmetry as extension of the Poincar\'e algebra a fully realised
supersymmetry algebra on the lattice must inevitably contain the generators of
translations which would imply arbitrary translations to be part of the
symmetry group of the lattice theory. By contrast, lattice regularised theories
are only symmetric under translations by the lattice spacing. Therefore a complete realisation of
the continuum supersymmetry algebra on the lattice is impossible and the full
supersymmetry can only be realised as an accidental symmetry in the continuum
limit of the lattice regularised theory. Technically, the reason for this can be
traced back to the failure of the Leibniz rule on the lattice \cite{Dondi:1976tx}.

It has been shown that even in supersymmetric quantum mechanics a naive
discretisation does not lead to a supersymmetric continuum limit
\cite{Giedt:2004qs}; generically, such a limit can at best be achieved by
fine-tuning the bare coefficients of all supersymmetry-breaking counterterms \cite{Montvay:1995rs}.
This, however, requires much knowledge of the theory in advance.
In some cases the relevant operators can be determined perturbatively,
cf.\ \cite{Golterman:1988ta}. To circumvent the fine-tuning process several
approaches are conceivable. Firstly a partial realisation of supersymmetry on the
lattice is possible for theories with extended supersymmetry (for a review see
\cite{Catterall:2009it}). Secondly recent developments aim at the construction
of a Ginsparg-Wilson inspired relation for supersymmetric theories to obtain a
lattice version of supersymmetry such that the continuum supersymmetry is
broken in a controlled way \cite{Bergner:2008ws}. Alternatively
for scalar theories a deformed supersymmetry algebra on the lattice can be
constructed by using a non-local product such that the theory is
invariant under the full (deformed) lattice supersymmetry
\cite{Bergner:2009vg,DAdda:2011jw}.

Apart from an explicit supersymmetry breaking by the finite lattice spacing there
exist further supersymmetry breaking effects that must be controlled in the
analysis of supersymmetric theories. For example, at finite temperature 
Lorentz invariance and therefore supersymmetry as extension of the Poincar\'e
symmetry are broken.\footnote{This problem can be avoided by choosing
periodic boundary conditions also for the fermions. However, this is only
possible for an unbroken supersymmetry.} In addition, for a finite spatial 
volume there may exist tunnelling processes between two formerly 
separate ground states such that the finite volume ground state energy 
is raised above zero. As it is inevitably to use finite lattices for 
numerical simulations these explicit supersymmetry breaking effects 
must be taken into account.

We begin our investigations with discretised supersymmetric quantum mechanics 
with dynamically broken supersymmetry in Sec.~\ref{chap:qm}. In 
this setting the basic concepts of
supersymmetric theories are explained and reference results for certain
observables are computed via the operator formalism, thus allowing to understand
the physics behind supersymmetry breaking on solid grounds. The corresponding
lattice regularisation is based on a formulation that has been used in the
unbroken supersymmetric quantum mechanics with great success
\cite{Bergner:2007pu}. We verify that even for quantum mechanical models with 
broken supersymmetry it is possible to obtain accurate results on the low 
lying energy spectrum from lattice simulations.

The minimal setting for a field theory with supersymmetry breaking phase
transition is given by the $\mathcal{N}=1$ Wess-Zumino model in $1+1$
dimensions, which is analysed in Sec.~\ref{chap:n1wz}. In the context of the
quenched model a particular renormalised critical coupling for the $\bbZ_2$
symmetry breaking is shown to be independent of the chosen lattice regulator.
The corresponding critical coupling in the full theory is determined and
the relation between $\bbZ_2$ and supersymmetry breaking is worked out.


\section{Broken supersymmetric quantum mechanics}
\label{chap:qm}

\noindent
An extensive analysis of quantum mechanical systems, such as the anharmonic
oscillator, with lattice regularised path integrals has been performed almost
three decades ago \cite{Creutz:1980gp}. There has been renewed interest in quantum
mechanical systems on the lattice in the context of supersymmetric quantum
mechanics (SQM). In several works SQM has been used as a toy model to study the
supersymmetry breaking induced by a naive lattice formulation
\cite{Catterall:2000rv} and to explore lattice regularisations with partially 
\cite{Beccaria:1998vi,Giedt:2004qs,Bergner:2007pu,Kastner:2007gz} or fully
\cite{Bergner:2009vg} conserved supersymmetries. It has been pointed out that
a discretisation without any conserved supersymmetries may not be free of
finite supersymmetry breaking renormalisation terms in the continuum limit
\cite{Giedt:2004vb} such that a careful treatment of supersymmetry restoration is
needed. Most of the lattice studies of SQM so far have been carried out for
the case of an unbroken supersymmetry and only few of them
\cite{Kanamori:2007ye,Kanamori:2007yx} consider the case of the dynamically
broken supersymmetry. Here, the case of a SQM with dynamically broken
supersymmetry is considered to explain the concepts and effects of supersymmetry
breaking in a setting that allows for high precision measurements in the
lattice theory and provides the possibility to compare to exactly calculable
reference values from the operator formalism.


\subsection{Operator formalism}
\label{sec:qm:opForm}

\noindent
SQM in one dimension is a generalisation of the
supersymmetric harmonic oscillator.\footnote{An extended introduction to the
operator formalism for supersymmetric quantum mechanics can be found in 
\cite{Cooper:1994eh,Wipf:2005sk}.} In analogy to supersymmetric field theories \emph{nilpotent}
supercharges $\sch$ and its adjoint $\sch^\dagger$ are introduced,
\begin{equation}
\sch^\dagger = \begin{pmatrix}
    0& 0\\ A& 0
    \end{pmatrix} = A \Psi^\dagger, \quad \sch = \begin{pmatrix}
                               0& A^\dagger\\ 0& 0
                               \end{pmatrix} = A^\dagger \Psi,
\end{equation}
with fermionic creation and annihilation operator $\Psi^\dagger$ and
$\Psi$ and first order differential operators
\begin{equation}
A = \frac{\dd}{\dd \phi}+P(\phi),\quad A^\dagger = -\frac{\dd}{\dd
\phi}+P(\phi)
\end{equation}
containing the \emph{prepotential} $P(\phi)$. In accordance to the field theory
language we denote the position operator of the quantum mechanical system
by $\phi$. The Hamiltonian is constructed via
\begin{equation}
\label{eq:qm:fullHamiltonian}
\frac{1}{2}\{\sch,\sch^\dagger\} =
\begin{pmatrix}
H_\text{B} &0\\ 0 & H_\text{F}
\end{pmatrix} = \frac{1}{2} \left(-\frac{\dd^2}{\dd \phi^2} +
P^2(\phi)+[\Psi^\dagger,\Psi]\, P'(\phi) \right)\equiv H,
\end{equation}
and acts on two-component state vectors $\ket{\psi} =
\left(\ket{\psi}_\text{B}, \ket{\psi}_\text{F}\right)^\trnsp$
where, for convenience, the first component is called
`bosonic' and the second one `fermionic'. The supersymmetry algebra is completed
by the nilpotency of $\sch$ and $\sch^\dagger$ and the commutation with $H$,
\begin{equation}
\{\sch,\sch\} = 0,\quad \{\sch^\dagger,\sch^\dagger \}=0,\quad [\sch,H] = 0.
\end{equation}
If $P$ is a linear function of $\phi$ then $A$ and
$A^\dagger$ are the bosonic annihilation and creation operators of the
(supersymmetric) harmonic oscillator. Accordingly the bosonic and fermionic
Hamiltonian is given by
\begin{equation}
\label{eq:qm:hamiltonians}
H_\text{B}= \frac{1}{2} A^\dagger A = -\frac{1}{2} \frac{\dd^2}{\dd \phi^2} +
V_\text{B} ,\quad H_\text{F}= \frac{1}{2}A A^\dagger =
-\frac{1}{2} \frac{\dd^2}{\dd \phi^2} +V_\text{F},\quad
V_{\text{B}/\text{F}} = \frac{1}{2}(P^2(\phi)\mp P'(\phi)).
\end{equation}
Both Hamiltonians are by construction non-negative. The bosonic
sector of a zero energy state is annihilated by $A$ and a fermionic one
is annihilated by $A^\dagger$,
\begin{equation}
H_\text{B}\ket{0}_\text{B} = 0 \;\Leftrightarrow\; A\ket{0}_\text{B} =
0, \quad H_\text{F}\ket{0}_\text{F} = 0 \;\Leftrightarrow\; A^\dagger
\ket{0}_\text{F} = 0.
\end{equation}
\begin{figure}
\centering
\includegraphics{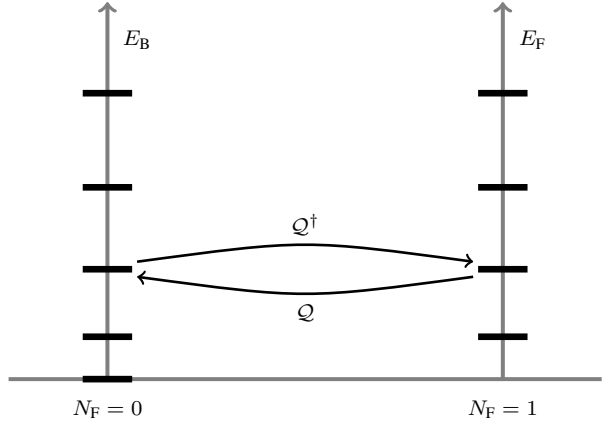}
\caption{\label{fig:qm:unbrokenSpectrum} Energy spectrum for the unbroken
supersymmetric quantum mechanics discussed in \cite{Bergner:2007pu}. $\sch$ and
$\sch^\dagger$ map between bosonic and fermionic sector.}
\end{figure}%
The supersymmetry algebra implies a strict pairing of excited
states, i.e.\ for every bosonic eigenstate 
$\ket{\psi_\text{B}}$ with energy $E>0$ there is a fermionic partner
state
\begin{equation}
\ket{\psi_\text{F}} = \frac{1}{\sqrt{2E}} \sch^\dagger\ket{\psi_\text{B}}, \quad
\ket{\psi_\text{B}} = \frac{1}{\sqrt{2E}} \sch\ket{\psi_\text{F}}
\end{equation}
with identical norm and energy.

The zero energy state(s) of the super Hamiltonian $H$ can be given explicitly
(in position space) as solutions of first order differential equations. If one of
these functions is normalisable, then the supersymmetric ground state exists and
supersymmetry is \emph{unbroken}. Since the product of possible zero energy
states $\braket{x}{0_\text{B}}\cdot \braket{x}{0_\text{F}}$ is constant, there
is at most \emph{one normalisable} state with zero energy. The explicit form 
of solutions implies that for a polynomial prepotential 
$P(\phi)=\sum_{n=0}^N c_n \phi^n$ with $c_N\neq 0$
supersymmetry is unbroken iff $N$ is odd. In that case there
is one normalisable zero energy state and the spectrum is similar to 
the one depicted in Fig.~\ref{fig:qm:unbrokenSpectrum}.


\subsubsection{Supersymmetry breaking and the Witten index}

\noindent
An existing and unbroken supersymmetry is defined by the existence of a
normalisable ground state $\ket{0}$ which is annihilated by $\sch$
and $\sch^\dagger$ which implies $H_\text{B}\ket{0}_\text{B}= H_\text{F}
\ket{0}_\text{F} = 0$. Witten introduced an \emph{index} \cite{Witten:1982df} to
determine whether supersymmetry can be broken dynamically in supersymmetric 
field theories. In the present context it is given by the trace over all eigenstates 
of $H$,
\begin{equation}
\Delta = \Tr(-1)^{N_\text{F}},
\end{equation}
where
$N_\text{F}=\begin{pmatrix}0 & 0\\ 0 & 1\end{pmatrix}$ is the \emph{fermion
number} operator that commutes with $H$.\footnote{As it stands, $\Delta$ is
not well defined and requires a normalisation, e.g.\
$\Delta=\lim_{\beta\to 0} \Tr[\ee^{-\beta H}(-1)^{N_\text{F}}]$.} Now, two
alternatives exist:
\begin{itemize}
  \item For \emph{broken} supersymmetry there is no normalisable zero energy
  state. All eigenstates of $H$ have positive energies and must be paired, which
  implies $\Delta = 0$.
  \item For \emph{unbroken} supersymmetry there are $n_\text{B}$ bosonic and
  $n_\text{F}$ fermionic ground states with zero energy. They contribute with
  $n_\text{B}-n_\text{F}$ to the Witten index. All contributions from the
  excited states cancel, which gives $\Delta = n_\text{B}-n_\text{F}$.
\end{itemize}
Therefore a non-vanishing Witten index implies an unbroken supersymmetry, but
not necessarily vice versa.
It is still possible that supersymmetry is unbroken while there are the same
number of bosonic and fermionic zero energy states. For a one dimensional
supersymmetric quantum mechanics at most one zero energy state is possible
and  $\Delta\neq 0$ is equivalent to unbroken supersymmetry.


\subsubsection{Specifying the model}
\label{ssec:qm:modelSpecs}

\noindent
The minimal modification of the supersymmetric harmonic oscillator with  broken
supersymmetry is given by the prepotential
\begin{equation}
P(\phi)=m\phi+h\phi^2
\end{equation}
with vanishing Witten index. Hence there is \emph{no
normalisable} ground state with zero energy. The spectrum is completely 
degenerate and acting with the supercharges on one
finite energy ground state will give the corresponding
superpartner of this ground state.\footnote{If not otherwise stated, ``ground
states'' may also have a positive energy.}

This model depends on the dimensionful parameters $m$ and $h$ and
in analogy to the supersymmetric harmonic oscillator $m$ is used
to set the scale. Therefore $f=h/m^{1.5}$ provides a scale independent
dimensionless coupling. In a heat bath
the dimensionless temperature is given by $T = (m \beta)^{-1}$ with $\beta$ as
dimensionful inverse temperature. Eventually coordinates are made dimensionless
by setting $\Phi=\phi\sqrt{m}$.

\begin{figure}
\centering
\includegraphics{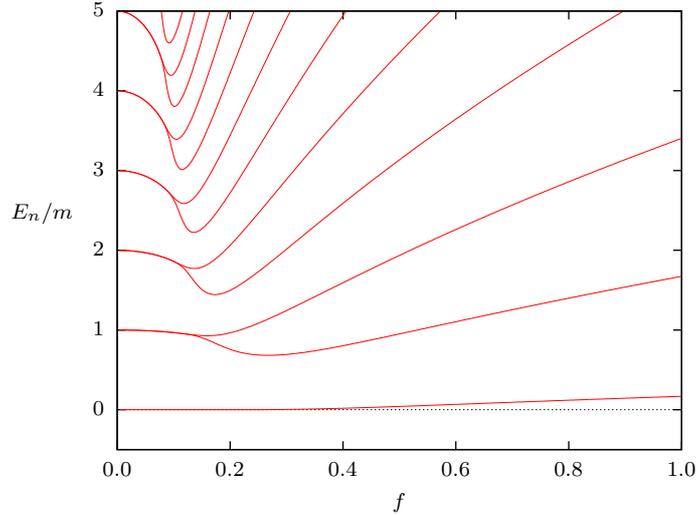}
\caption{\label{fig:qm:hamiltonianSpectum} Energy levels for the broken
supersymmetric quantum mechanics. Each level is doubly degenerate with one
bosonic and one fermionic eigenstate.}
\end{figure}%
The energy spectrum and corresponding states can be computed directly 
by a discretisation of position space and replacement of $\frac{\dd}{\dd
\phi}$ by a discretised derivative. After the analysis of different possible
discretisations in  \cite{Kirchberg:2004vm} the numerically most stable choice
is the SLAC derivative \cite{Drell:1976mj}, which for periodic
boundary conditions on a lattice with an odd number of points $N$ and lattice
spacing $a$ is given by
\begin{equation}
\label{eq:qmb:slac}
\begin{aligned}
a(\partial^\text{SLAC})_{xy} &= \begin{cases}
                              0 &\hskip3.3mm\colon x=y\\
                              \frac{\pi}{N}(-)^{x-y} \frac{1}{\sin (\pi
                              (x-y)/(Na))} &\hskip3.3mm \colon x\neq y
                              \end{cases},\\
-a^2(\partial^\text{SLAC})^2_{xy} &= \begin{cases}
                              \frac{\pi^2}{N^2}\frac{N^2-1}{3} &\colon x=y\\
                              \frac{2\pi^2}{N^2}(-)^{x-y} \frac{\cos (\pi
                              (x-y)/(Na))}{\sin^2 (\pi
                              (x-y)/(Na))} & \colon x\neq y
                              \end{cases}.
\end{aligned}
\end{equation}

\begin{figure}
\centering
\includegraphics{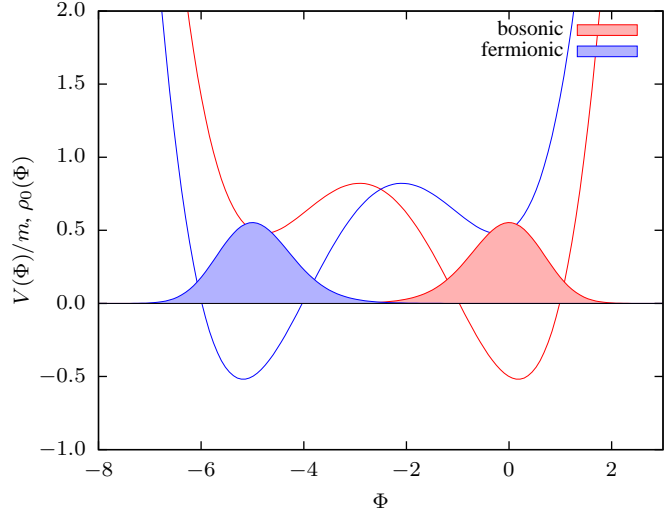}
\caption{\label{fig:qm:groundStates02} Probability density
$\rho_0(\Phi)$ (shaded areas) for bosonic and fermionic ground
state and potentials $V_{\text{B}/\text{F}}$ (lines) of the corresponding
Hamiltonian at coupling $f=0.2$.}
\end{figure}

The spectrum of the diagonalised Hamiltonian is depicted in
Fig.~\ref{fig:qm:hamiltonianSpectum}. For weak couplings $f\lesssim 0.1$ there is an
additional (approximate) degeneracy of the excited spectrum corresponding to
the perturbed energy levels of two harmonic oscillators with energies
$\mathbb{N} m$ residing at the minima of the bosonic and fermionic potential
$V_{\text{B}/\text{F}}$ (see Fig.~\ref{fig:qm:groundStates02}).


\paragraph{Interpretation as a physical system}

The naming `bosonic' and `fermionic' sector may sound misleading because of the
complete degeneracy of the spectrum. The system can be interpreted as a
particle with spin $1/2$ moving in an external potential that depends on the spin
orientation. So `bosonic' may refer to `spin down' and `fermionic' to `spin up',
respectively. Supersymmetry in this case is represented as degeneracy between an
up and a down state. For the case of unbroken supersymmetry the ground state is
unique and is invariant under application of supersymmetry although it is in a
definite spin state given by the interaction potential. For the broken
supersymmetry there are (in the present case) two different ground states none
of which is energetically preferred (see Fig.~\ref{fig:qm:groundStates02}).
There is no interaction given by the Hamiltonian between bosonic and fermionic sector and one ground state  will be preserved if
no external interaction is applied (e.g.\ by interacting with a heat bath at
finite temperature). Applying the supercharge will 
give the partner ground state and amounts to the symmetry between spin up and
spin down state. Furthermore no linear combination of the two ground states is
invariant under the supersymmetry.


\paragraph{Physics at $T=0$}

At vanishing temperature physics is given by ground state (vacuum)
expectation values. Since supersymmetry is broken the system will
stay in one of the degenerate ground states and expectation values are
defined by this particular ground state.\footnote{This is similar to
the $\bbZ_2$ symmetry in the Ising chain. There, at any finite
temperature the symmetry is restored. Only for
$T=0$ the system will take (and preserve) one of the possible ``ground
states''.} Without loss of generality results are given for the bosonic (finite
energy) ground state $\ket{0_\text{B}}$ and the expectation value of an
observable $\mathcal{O}$ is thus given by $\vev{\mathcal{O}}_0 =
\bra{0_\text{B}}\mathcal{O}\ket{0_\text{B}}$.

Observables
can be computed from the diagonalised Hamiltonian to provide reference values
for lattice computations. This is done in the next section. On the lattice the primary focus lies on one- and two-point functions and the probability density of the
coordinate $\Phi$ given by $\rho_0(\Phi) = \abs{\braket{\Phi}{0}}^2$. The one-point
function is then given by $\vev{\Phi}_0 = \int \dd \Phi\, \rho_0(\Phi) \Phi$.

The bosonic two-point function (in the bosonic ground state) is defined
through the Euclidean time evolution,
\begin{equation}
\vev{\Phi(t)\Phi(0)}_0 = \bra{0_\text{B}} \Phi(t) \Phi(0)\ket{0_\text{B}}
= \bra{0_\text{B}} \ee^{tH} \Phi \ee^{-tH} \Phi\ket{0_\text{B}}
= \bra{0_\text{B}} \Phi \ee^{-t(H-E_0)} \Phi\ket{0_\text{B}}.
\end{equation}
Equivalently the fermionic correlation function is computed by
\begin{equation}
\vev{\Psi(t)\Psi^\dagger(0)}_0 
= \bra{0_\text{B}} \Psi \ee^{-t(H-E_0)} \Psi^\dagger \ket{0_\text{B}}.
\end{equation}
In each case $t m$ defines the dimensionless `time'.

The last quantity of interest is the effective potential which may be either
defined by a Legendre transform of the Schwinger function\footnote{The
Schwinger function is naturally defined in a  path integral formulation.}
or more directly at vanishing temperature by
\begin{equation}
\label{eq:qm:effectivePot}
V_\text{eff}(\Phi_0) = \min_{\bra{\psi}\Phi\ket{\psi}=\Phi_0}\bra{\psi} H
\ket{\psi}.
\end{equation}


\paragraph{Finite temperature physics}

For any finite temperature there is a Boltzmann distribution 
with the same contribution of bosonic and fermionic partner states, including
the lowest energy ground states. Again, with high precision calculations of
the low lying spectrum provided by the diagonalised Hamiltonian it is possible
to compute the thermal field distribution and expectation values
\begin{equation}\label{eq:qm:singelSiteThermal}
\rho_T(\Phi) = Z^{-1} \sum_{E} \ee^{-E/T}
\abs{\braket{\Phi}{\psi_{E}}}^2,\quad
\vev{\mathcal{O}}_T = Z^{-1} \sum_{E} \bra{\psi_{E}} \ee^{-\beta H} \mathcal{O}
\ket{\psi_{E}}, \quad Z = \sum_{E} \ee^{-E/T},
\end{equation}
where the sums run over all bosonic and fermionic states.


\subsection{Lattice regularised path integral}
\label{sec:qm:lat}

\noindent
With the methods given in the previous section it is possible to obtain exact
results against which the path integral based calculations can be compared.
Therefore the accuracy of the lattice simulation can be determined
even when supersymmetry is broken. The corresponding Euclidean path
integral is given by
\begin{equation}
\mathcal{Z} = \int \Dd \phi\, \Dd \psi\, \Dd \bar\psi\,
\ee^{-S[\phi,\psi,\bar\psi]},
\end{equation}
with Euclidean action
\begin{equation}
S = \int \dd \tau\,\left(\frac{1}{2}(\partial \phi)^2 +
\frac{1}{2}P^2(\phi) + \bar\psi (\partial + P'(\phi)) \psi\right).
\end{equation}
Expectation values are computed via
\begin{equation}
\vev{A} = \mathcal{Z}^{-1}  \int \Dd \phi\, \Dd \psi\, \Dd \bar\psi\,
A[\phi,\psi,\bar\psi] \ee^{-S[\phi,\psi,\bar\psi]}.
\end{equation}
Supersymmetry appears as a symmetry of the action, where one transformation is
given by
\begin{equation}
\delta^{(1)} \phi = \bar\varepsilon \psi,\quad \delta^{(1)} \bar\psi = -\bar
\varepsilon (\dot \phi+P(\phi)),\quad \delta^{(1)}\psi = 0
\end{equation}
and a variation of the action gives 
$\delta^{(1)} S = \int \dd \tau \left[\partial (\bar\varepsilon P\psi)\right] =
0$.
In the same way the action allows for a second supersymmetry transformation
\begin{equation}
\delta^{(2)} \phi = \bar\psi \varepsilon,\quad \delta^{(2)} \bar\psi = 0,\quad
\delta^{(2)}\psi = (\dot\phi-P)\varepsilon.
\end{equation}
For the above supersymmetries to hold it is necessary that the fields vanish at
infinity or that they are periodic in the Euclidean time. But for a
thermal path integral at inverse temperature $\beta$ the
fields obey the boundary conditions
\begin{equation}
\phi(0) = \phi(\beta),\quad \psi(0) = -\psi(\beta),\quad \bar\psi(0) =
-\bar\psi(\beta),
\end{equation}
i.e.\ the fermionic field is \emph{antiperiodic} in time.
Since the fields need not vanish anymore the variation of the action then reads
\begin{equation}
\delta^{(1)} S = \left[\bar\varepsilon P\psi\right]_{\tau=0}^\beta =
-2\left[\bar\varepsilon P\psi\right]_{\tau=0}
\end{equation}
which can be non-vanishing so that supersymmetry is
broken by the finite temperature. In \cite{Catterall:2000rv,Bergner:2007pu}
for an unbroken supersymmetric quantum mechanics periodic boundary conditions
have been used to avoid such an explicit breaking. For temperature going to zero
a change in boundary conditions is equivalent to an insertion of $(-1)^{N_\text{F}}$ into the path integral,
\begin{equation}
\mathcal{Z}_\text{p} = \int \Dd \phi\, \Dd \psi_\text{p}\, \Dd
\bar\psi_\text{p}\, \ee^{-S[\phi,\psi,\bar\psi]} = \int \Dd \phi\, \Dd
\psi_\text{ap}\, \Dd \bar\psi_\text{ap}\,
(-1)^{N_\text{F}}\ee^{-S[\phi,\psi,\bar\psi]} = \mathcal{Z}_\text{ap}
\Delta.
\end{equation}
Here, the periodic path integral is vanishing due to $\Delta=0$ for a broken
supersymmetry. Thus, for a theory allowing 
supersymmetry breaking, periodic (supersymmetry preserving) boundary conditions
cause a severe sign problem. This does not completely rule out the choice of these boundary
conditions, as will be discussed on the case of the two dimensional
$\mathcal{N}=1$ Wess-Zumino model in Sec.~\ref{sec:n1:full}, but puts
constraints on the range of applicability.\footnote{For a phase with unbroken supersymmetry in models with
$\Delta=0$ only one specific ground state belongs to the physical spectrum and
periodic boundary conditions may be imposed.} To have a well defined
(non-vanishing) path integral antiperiodic (thermal) boundary conditions for
the fermionic fields are imposed.

For a construction of a lattice model the choice of the
lattice regularised derivative is crucial.\footnote{In contrast to the operator
formalism where the field space is discretised, the lattice path integral is
based on a discretisation in the Euclidean time.} The canonical choice for
scalar theories would be the forward (or equivalently backward) derivative.
For derivatives appearing in the fermionic action a popular choice is given by
Wilson's prescription \cite{Wilson:1975id}. Nevertheless, these simple
discretisation rules are not applicable to supersymmetric theories as analysed in
\cite{Catterall:2000rv,Bergner:2007pu,Giedt:2004vb} for the case of an unbroken
supersymmetric quantum mechanics. These results show the need for a more careful
treatment of the discretisation of supersymmetric theories. In the comparative
study of six different discretisations \cite{Bergner:2007pu} the one
based on the SLAC derivative provides results close to the continuum limit even
at finite lattice spacing. For an odd number of lattice points with periodic boundary conditions
the matrix representation is already given in Eq.~\eqref{eq:qmb:slac}.
Antiperiodic boundary conditions (necessary for fermionic fields) are best
realised on an even lattice with $N$ points,
\begin{equation}
a(\partial^\text{SLAC})_{xy} = \begin{cases}
                              0 &\colon x=y\\
                              \frac{\pi}{N}(-)^{(x-y)/a} \frac{1}{\sin (\pi
                              (x-y)/(Na))} & \colon x\neq y
                              \end{cases},
\end{equation}
while the squared SLAC derivative for an even number of lattice points and
periodic boundary conditions (as needed for the bosonic fields) reads 
\begin{equation}
-a^2(\partial^\text{SLAC})^2_{xy} = \begin{cases}
                              \frac{\pi^2}{N^2}\frac{N^2+2}{3} &\colon x=y\\
                              \frac{2\pi^2}{N^2}(-)^{(x-y)/a} \frac{1}{\sin^2
                              (\pi (x-y)/(Na))} & \colon x\neq y
                              \end{cases}.
\end{equation}
Although it was analysed \cite{Karsten:1979wh} that this prescription will lead
to a non-covariant and non-local continuum limit in lattice QED it can be
proven \cite{Bergner:2007pu,bergner:2009phd} that for scalar theories in one
or two dimensions with Yukawa interactions a local renormalisable continuum
limit is reached. For that reason the SLAC derivative is used here to
regularise the supersymmetric quantum mechanics on the lattice with
corresponding action
\begin{equation}
S = -\sum_{x,y} \half\hat\phi_x(\hat\partial^\text{SLAC})^2_{xy}\hat\phi_y
+ \half \sum_x P(\hat\phi_x)^2 + \sum_{x,y}
\bar\psi_x\bigl(\hat\partial^\text{SLAC}_{xy}+P'(\hat\phi_x)\delta_{xy}\bigr)\psi_y
\end{equation}
on lattices with an even number of sites, where field $\hat\phi$ and
derivative $\hat\partial^\text{SLAC}$ are dimensionless and arise from a
rescaling of the dimensionful quantities with the lattice spacing.\footnote{The fermionic
fields $\psi$, $\bar\psi$ are already dimensionless and do not need to be
rescaled.}

\begin{figure}
\hfill
\includegraphics{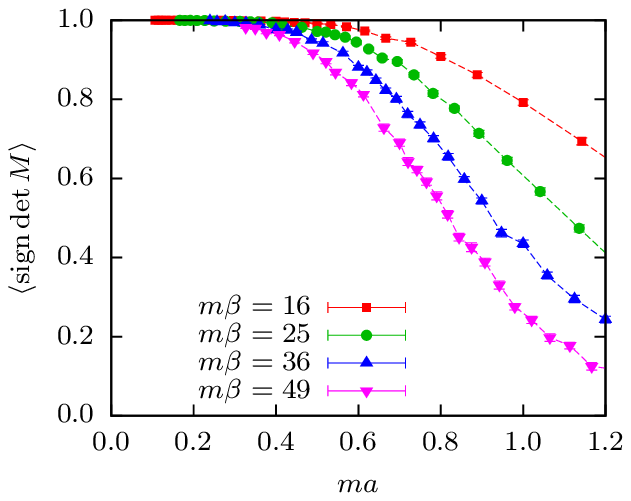}
\hfill
\includegraphics{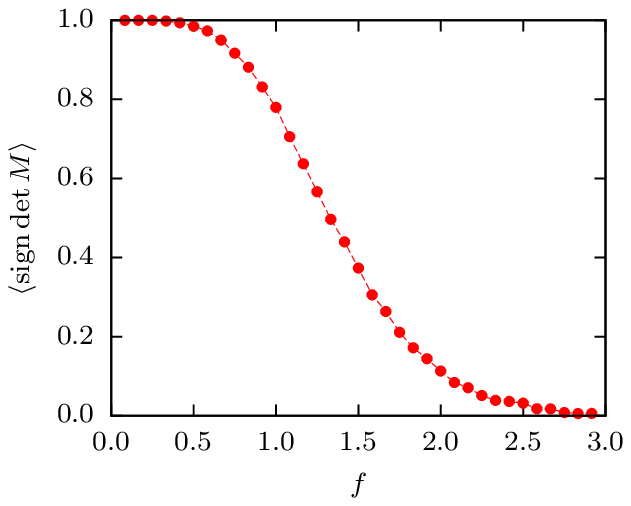}
\hfill\hfill\phantom.
\caption{\label{fig:qm:signDeterminant} Sign of the fermion determinant measured
at fixed $f=1$ (left panel, $10^6$ configurations per data point) and at fixed
$m\beta=36$ and $N=50$ lattice points (right panel, $5\cdot 10^6$
configurations per data point).}
\end{figure}%
Access to the non-perturbative sector of the lattice model is gained from
Monte-Carlo simulations which have become a powerful tool due to
increasing computer power and algorithmic improvements that allow for the
inclusion of dynamical fermions in simulations. These statistical methods are
based on importance sampling and the interpretation of the lattice regularised
path integral
\begin{equation}
\mathcal{Z} = \int \DD \hat\phi\, \DD \psi\, \DD \bar\psi\,
\ee^{-S[\hat\phi,\psi,\bar\psi]}
\end{equation}
as probability distribution. To construct the probability density the
action is then split into a bosonic and fermionic part according to
\begin{equation}
S[\hat\phi,\psi,\bar\psi] = S_\text{B}[\hat\phi] +
\sum_{x,y}\bar\psi_x M_{xy}[\hat\phi]\psi_y.
\end{equation}
Applying the rules of Grassmann integration the fermionic part
of the path integral can be integrated out and yields
\begin{equation}
\mathcal{Z} = \int \DD \hat\phi\, \det M[\hat\phi]\,
\ee^{-S_\text{B}[\hat\phi]}.
\end{equation}
In this way (bosonic) expectation values are computed by
\begin{equation}
\vev{\mathcal{O}[\hat\phi]} = \mathcal{Z}^{-1} \int \DD \hat\phi\,
\mathcal{O}[\hat\phi]\,\det M[\hat\phi]\, \ee^{-\SB[\hat\phi]}.
\end{equation}
In a Monte-Carlo simulation the lattice regularised fields $\hat\phi$ are
generated according to the distribution
\begin{equation}
\label{eq:qm:probabilityDistribution}
\rho[\hat\phi] = \ee^{-S_\text{B}[\hat\phi] + \ln \abs{\det M[\hat\phi]}}.
\end{equation}
After a number of $N_\text{MC}$ samples one obtains a time series
$\hat\phi^{(k)}$, $k=1,\ldots,N_\text{MC}$, and expectation values are evaluated
using
\begin{equation}
\vev{\mathcal{O}} \overset{N_\text{MC}\to\infty}{=}
N_\text{MC}^{-1}\sum_{k=1}^{N_\text{MC}} \mathcal{O}[\hat\phi^{(k)}].
\end{equation}
This expression is only exact iff $\det M\geq 0$. If $\det M$ is
negative the sign has to be taken into account by reweighing. However, the
emphasis shall lie on the physical questions and for further simulation details
the reader is referred to the rich literature on Monte-Carlo methods, e.g.
\cite{newman,binder,Clark:2006fx,Frezzotti:1997ym,Kastner:2008zc,Toral:1994zt}.


\subsubsection{Sign of the fermion determinant}
\label{ssec:qm:sgnDet}

\noindent
For periodic fermionic boundary conditions $\ZZ_\text{p} \propto
\Delta$ will vanish in the continuum and a severe sign problem is expected to
arise in reweighed expectation values. For thermal boundary conditions it is a
priori unknown if there are configurations with $\det M <0$ and if there is any dependence on lattice spacing, temperature, or
coupling. For that reason $\vev{\sign \det M}$ has been measured in the sign
quenched ensemble with the distribution given by
Eq.~\eqref{eq:qm:probabilityDistribution} for different parameter sets (see Fig.~\ref{fig:qm:signDeterminant}). These results imply a complete absence of
the sign problem in the continuum limit for \emph{every} coupling and
temperature. The sign problem only exists for large couplings $f$ 
at \emph{fixed} lattice spacing and temperature.


\subsubsection{Ground state structure}
\label{ssec:qm:groundStateStruct}

\begin{figure}
\hfill
\includegraphics{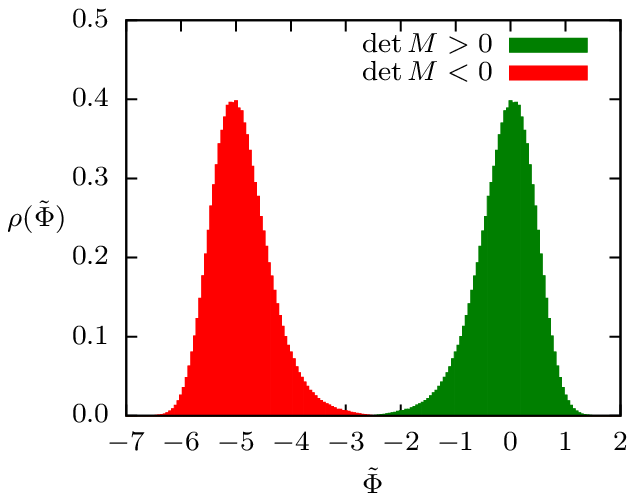}
\hfill
\includegraphics{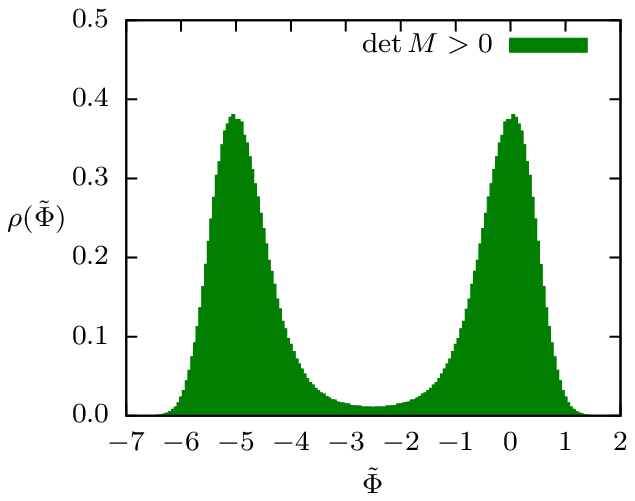}
\hfill\hfill\phantom{.}
\caption{\label{fig:qm:boundaryCond} The distribution of the averaged field
$\tilde\Phi$ for $m\beta=4$ at coupling $f=0.2$ with respect to the sign
of the determinant for periodic (left panel, $N=101$) and antiperiodic (right
panel, $N=100$) fermionic boundary conditions obtained from $10^6$
configurations.}
\end{figure}%
\noindent
With thermal as well as supersymmetry preserving (periodic) boundary conditions
for low temperature $T=0.25$ the ground state structure is
analysed.\footnote{Periodic boundary conditions have only been used for this
particular analysis of the ground state structure.} Simulations at $f=0.2$ are
performed and the distribution of the lattice averaged field $\tilde\Phi = N^{-1}\sum_x \Phi_x$
is analysed with respect to the sign of $\det M$ (see
Fig.~\ref{fig:qm:boundaryCond}). Configurations with $\tilde\Phi>-1/(2f)$ are 
unaffected by a change of boundary conditions whereas the sign of $\det M$
changes for $\tilde\Phi<-1/(2f)$. This behaviour can be seen explicitly
on the level of the discretised action. For the chosen prepotential bosonic and
fermionic ground state are related by a $\bbZ_2$ symmetry $\Phi_x\to
-\Phi_x-1/f$. $\SB$ is invariant under the symmetry operation whereas
the effect on the fermionic contribution depends on the derivative used. The
SLAC derivative has an antisymmetric matrix representation,
$\partial^\text{SLAC}_{xy} = -\partial^\text{SLAC}_{yx}$. $P'(\phi)$ enters on
the diagonal of the fermion matrix $M$.
Applying the $\bbZ_2$ symmetry gives
$P'(\phi)\to -P'(\phi)$ and changes the sign of the diagonal elements
of the fermion matrix. Altogether, the symmetry operation changes
$M(\hat\phi)\to -M^\trnsp(\hat\phi)$. For antiperiodic (periodic) fermions the
fermion matrix size will be even (odd, respectively) and the determinant will keep the
modulus but changes its sign for periodic boundary conditions while for
antiperiodic fermions the sign is preserved. Therefore periodic SLAC fermions
imply $\ZZ_\text{p} = 0$ exactly. The boundary condition
dependence of the distribution coincides with introducing $(-1)^{N_\text{F}}$ into the path
integral for periodic boundary conditions and configurations with
$\tilde\Phi>-1/(2f)$ correspond to the bosonic ground state whereas the
other ones correspond to the fermionic ground state, respectively. This is in
accordance with results from the operator formalism shown in
Fig.~\ref{fig:qm:groundStates02}.


\subsubsection{Thermal field distribution}

\begin{figure}
\hfill
\includegraphics{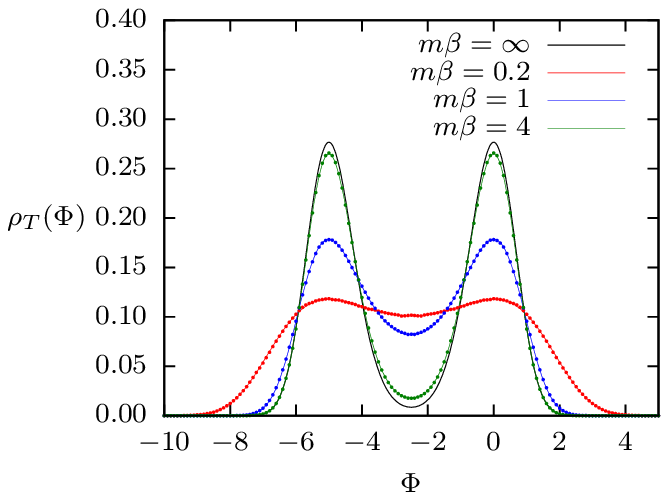}
\hfill
\includegraphics{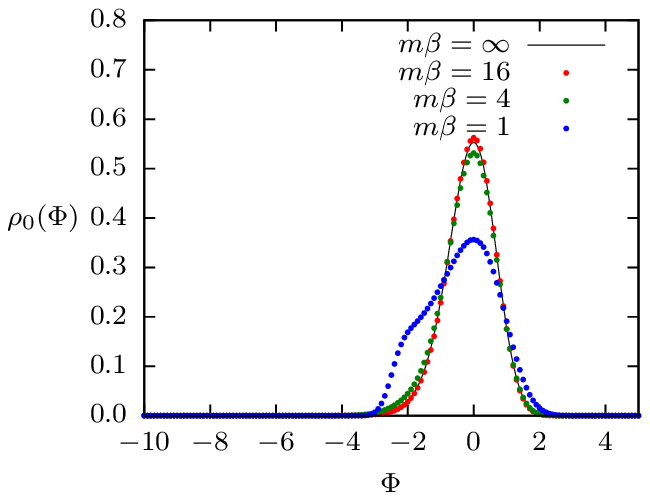}
\hfill\hfill\phantom.
\caption{\label{fig:qm:thermalDistribution}Probability distribution $\rho(\Phi)$
for different temperatures at coupling $f=0.2$ on a $N=100$ lattice.
Lines depict the exact results, points arise from the lattice calculations. For
comparison the distribution at $T\to 0$ is also drawn. Left panel: Thermal
distribution. Right panel: Exact distribution for the bosonic ground state and
distribution measured on configurations with $\tilde\Phi>-\frac{1}{2f}$.}
\end{figure}%
\noindent
At finite temperature the single site distribution $\rho_T(\Phi)$ of
Eq.~\eqref{eq:qm:singelSiteThermal} is computed on a lattice with $N=100$ points
at fixed coupling $f=0.2$ in the temperature range $m\beta\in [0.2,4]$ with
$10^6$ configurations. Even for this coarse lattice the results match
almost perfectly with the reference values from the diagonalised Hamiltonian 
(see Fig.~\ref{fig:qm:thermalDistribution}, left panel). At low temperature 
$m\beta=16$ it is possible to extract the probability
distribution in the bosonic ground state by considering only
configurations with $\tilde\Phi>-1/(2f)$. Even at finite (but small) temperature
the so-extracted probability distribution follows the exact result at $T=0$, see 
Fig.~\ref{fig:qm:thermalDistribution} (right panel). In consequence, for $T\to
0$ a thermal mixture of bosonic and fermionic ground state is found whereas at exactly vanishing temperature the system can be triggered to stay in one ground state.


\subsubsection{Effective potential}

\noindent
The effective potential as introduced in Sec.~\ref{ssec:qm:modelSpecs} is not
directly accessible in lattice simulations. A better suited quantity is given by the
constraint effective potential \cite{ORaifeartaigh:1986hi}
\begin{equation}
U(\tilde\Phi_0) = -\frac{1}{\beta} \ln \left( \int \DD \hat\phi\, \det
M[\hat\phi]\, \ee^{-\SB[\hat\phi]}\, \delta(\tilde\Phi-\tilde\Phi_0)\right),
\end{equation}
which can be straightforwardly computed on the lattice.\footnote{Only an
additive normalisation constant of the constraint effective potential is left
undetermined.} It has been proven for the case of bosonic scalar theories that
the constraint effective potential coincides in the 
limit of infinite spacetime volume (or here equivalently,
vanishing temperature) with the effective potential,
$\lim_{m\beta\to\infty} U(\tilde\Phi) = \lim_{m\beta\to\infty}
V_\text{eff}(\tilde\Phi)$.

\begin{figure}
\centering
\includegraphics{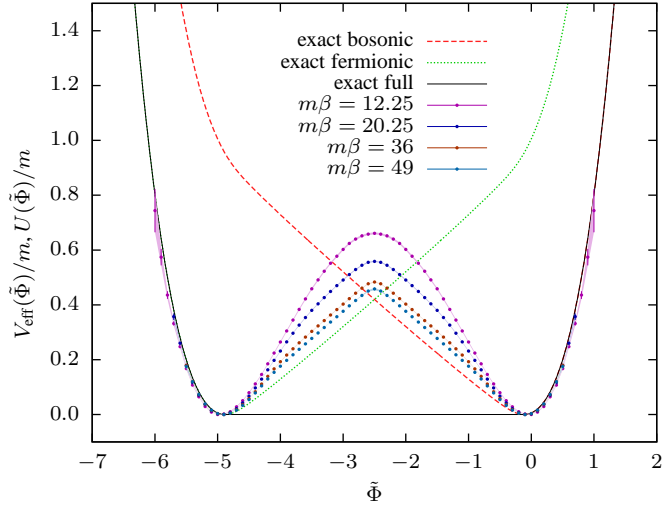}
\caption{\label{fig:qm:effPot} Effective potential for the full model
("exact full"), bosonic sector ("exact bosonic"), and fermionic sector ("exact
fermionic") of the Hamiltonian at $T=0$ and constraint effective
potential from lattice simulations of the full model at various $m\beta$ for
coupling $f=0.2$ and $N=300$ lattice points, measured with up to $10^7$
configurations.}
\end{figure}

We determined the constraint effective potential for various 
inverse temperatures $m\beta$ at fixed coupling $f=0.2$ for $N=300$ 
lattice points. The result is depicted in Fig.~\ref{fig:qm:effPot} together with the
effective potential for the full theory and for the bosonic/fermionic sector. To
avoid the ambiguities of the unknown additive constant, the minimum of each
potential is normalised to $0$.\footnote{This amounts to an offset for the
effective potential of the full theory that is given by the ground state energy,
$\Delta V_\text{eff}= E_0 = 3.68\cdot 10^{-5} m$.}

It is often stated that the conventional effective potential for a
quantum mechanical system is strictly convex. This is true for systems 
with Hilbert space $L^2(\mathbb{R}^d)$. For the supersymmetric
system the Hilbert space is $L^2(\mathbb{R})\otimes \mathbb{C}^2$ and all linear
combination of the bosonic and fermionic ground states have the same energy, which
implies a flat region in the effective potential as defined in
Eq.~\eqref{eq:qm:effectivePot}.

To compute the value of the constraint effective potential even around the
peak $\tilde\Phi=-1/(2f)$, reweighting methods \cite{Ferrenberg:1988yz} similar to the
multicanonical ensemble \cite{Berg:1991cf} have been applied. Simulations were
performed with a modified action $S_B' = S_B-W(\tilde\Phi)$, with
$W(\tilde\Phi)$ suitably chosen by iterative refinement to approximate
$\beta U(\tilde\Phi)$ between the two minima and $W(\tilde\Phi)$ constant in the
outer regions. Of course, the configurations must be (re-)weighted after simulation with a factor
$\exp(-W(\tilde\Phi))$.\footnote{In a conventional Monte-Carlo simulation
without reweighting the configurations around the peak will be suppressed at $m\beta=49$ by more than $\ee^{-20}$.}

It is apparent that the positions of the minima
of $U(\tilde\Phi)$ correspond to the bosonic and fermionic ground state.
However, the constraint effective potential shows no tendency to flatten out
towards the conventional effective potential. It rather tends towards
\begin{equation}
\lim_{m\beta\to\infty} U(\tilde\Phi) = \lim_{m\beta\to\infty}
\min\left(V_{\text{eff},\text{B}}(\tilde\Phi),V_{\text{eff},\text{F}}(\tilde\Phi)\right),
\end{equation}
where $V_{\text{eff},\text{B}/\text{F}}$ denotes the effective potential of the
bosonic (fermionic, respectively) sector of the Hamiltonian
\eqref{eq:qm:fullHamiltonian}.

This result can be explained by the behaviour of fermions on the
lattice under a change of boundary conditions. The ratio of the fermion
determinant between periodic and antiperiodic boundary conditions can be
computed in the continuum limit
\cite{Gozzi:1984jx}, and by using zeta function regularisation
\cite{BoschiFilho:1995ws} one gets rid of the free determinant as normalisation
constant. The continuum result
\begin{equation}
R_\beta[\phi]\equiv\frac{\det M_\text{p}[\phi]}{\det M_\text{ap}[\phi]} =
\frac{\sinh\bigl[\half \int_0^\beta
P'(\phi(\tau))\dd\tau\bigr]}{\cosh\bigl[\half \int_0^\beta
P'(\phi(\tau))\dd\tau\bigr]}
\end{equation}
will be either $+1$ or $-1$ in the infinite volume limit, depending on the
value of $\tilde\Phi$,
\begin{equation}
\lim_{\beta\to\infty} R_\beta(\phi) = \begin{cases}
+1&\colon \tilde\Phi>-\half f^{-1}\\
-1&\colon \tilde\Phi<-\half f^{-1}
\end{cases},
\end{equation}
which is equivalent to $(-1)^{N_\text{F}}$.
I.e by means of the lattice path integral with
fixed (thermal) boundary conditions it is only possible to assess either the
zero fermion sector or the one fermion sector. The interpolating states
between both sectors that are responsible for the flattening of the effective
potential are thus not accessible by the (lattice) path integral at any finite
temperature.


\subsubsection{Two-point functions and spectrum}

\begin{figure}
\hfill
\includegraphics{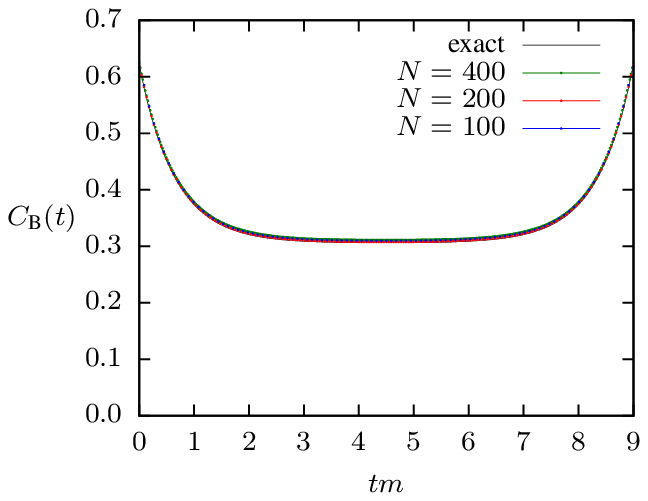}
\hfill
\includegraphics{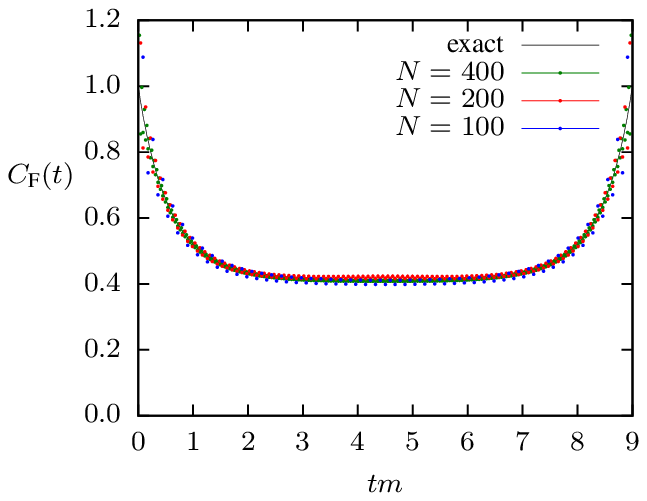}
\hfill\hfill\phantom.
\caption{\label{fig:qm:thermalCorrelators} The two-point function
(left panel: bosonic, right panel: fermionic) for the thermal ensemble given by
$m\beta=9$ and coupling $f=1$ computed by the diagonalised Hamiltonian and on
a lattices with $N\in\{100,200,400\}$ points. For the bosonic case the data
points coincide with the reference values. A statistics of up to $10^7$
configurations has been used.}
\end{figure}%
\noindent
Lattice based path integral methods provide a non-perturbative way to gain
information about the spectrum of the theory. Via the long distance behaviour
of correlators $C(t)\xrightarrow{t\to \infty}
\exp(-m_\text{phys}t)$ it is possible to extract the physical `pole mass'
$m_\text{phys}$ which is given by the imaginary part of the pole of the
propagator $G(p)=(\mathcal{F}C)(p)$, the Fourier transform of the correlator,
and describes the energy difference between the ground state and first excited state
of the theory. For that reason connected correlation functions in the thermal ensemble have been computed for
bosonic and fermionic fields (see Fig.~\ref{fig:qm:thermalCorrelators}) with
\begin{equation}
C_\text{B}(t) = \vev{\Phi(t)\Phi(0)}-\vev{\Phi}^2\quad\text{and}\quad
C_\text{F}(t) = \vev{\psi(t)\bar\psi(0)}.
\end{equation}
The correlators take non-vanishing constant values for large distances in a
region where the exponential falloff drops below the visibility
scale.\footnote{This is \emph{not directly} related to any unconnected part.
Here only connected correlators are considered.} Correlators computed 
from the lattice regularised theory fit nicely to the ones computed by the diagonalised Hamiltonian.
Fluctuation are still visible around the continuum values with the size of
fluctuations vanishing for smaller lattice spacings. Further it is possible
from the fermionic correlator at small (but non-vanishing) temperature to
compute the overlap of bosonic and fermionic ground state
$\abs{\bra{0_\text{F}}\Psi^\dagger\ket{0_\text{B}}}^2=0.41174$ by the
approximation $C_\text{F}(m\beta/2)=0.419(7)$ for $m\beta=16$ and $N=400$.

\begin{figure}\hfill
\includegraphics{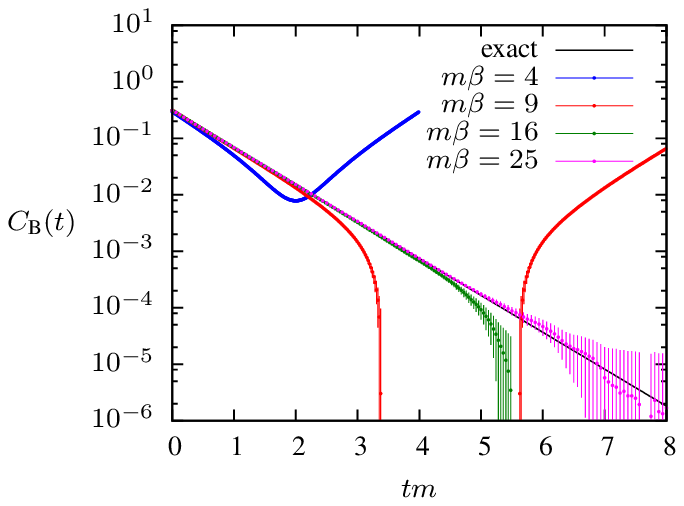}
\hfill
\includegraphics{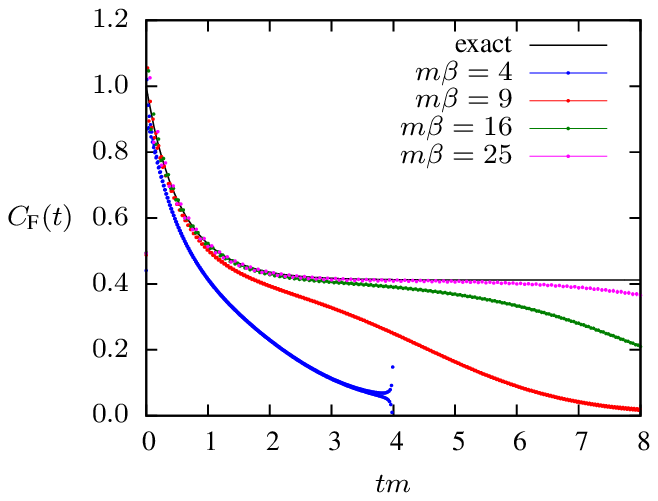}
\hfill\hfill\phantom.
\caption{\label{fig:qm:groundStateCorrelators} Bosonic (left
panel) and fermionic (right panel) correlator in the ensemble projected to one
ground state as obtained from the diagonalised Hamiltonian and from lattice
simulations with $N=400$ sites at coupling $f=1$ and statistics of up to $3\cdot
10^7$ configurations. }
\end{figure}%
With nearly vanishing temperature the system will mainly reside in the ground
states, and with the results of Sec.~\ref{ssec:qm:groundStateStruct} it is
possible to compute the correlation function at $T=0$ by projecting
to one of the ground states (see Fig.~\ref{fig:qm:groundStateCorrelators}). In that case
the bosonic correlator shows no constant
part and the exponential behaviour completely coincides with the one resulting from the
first excited (bosonic) state. From the lattice values at $N=400$ a mass of
$m_\text{phys}=1.5047(15)\,m$ is extracted through an exponential fit in the
range $tm\in[1,3]$ from a simulation at $m\beta=25$. This is very close
to the exact value $E_1-E_0=1.5046\,m$. Of course this method is only
applicable for $t<\beta/2$ and there will be large deviations for $t$ close to
$\beta/2$.

\begin{figure}
\centering
\includegraphics{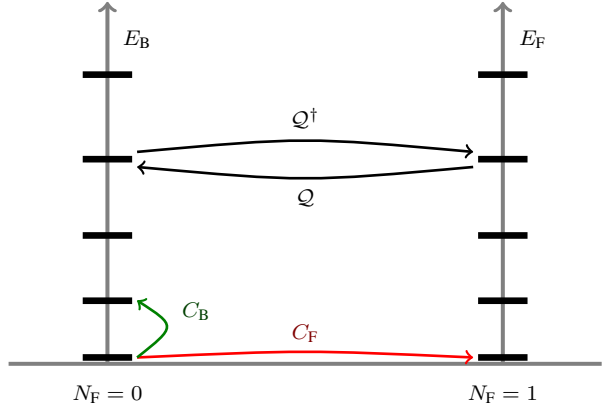}
\caption{\label{fig:qm:groundStateExcitations} Excitations visible as $t\to
\infty$ behaviour of the correlators evaluated on configurations projected to
the bosonic ground state.}
\end{figure}%
All these results clearly demonstrate that it is possible to extract correlators at
finite and zero temperature from a lattice discretisation in complete
agreement with exact results in the continuum limit. 
The degenerate ground states are visible as a constant part in the fermionic
correlator and a projection to one ground state allows to extract the energy
difference between the ground state and first excited state (see Fig.~\ref{fig:qm:groundStateExcitations}).


\subsubsection{Ward identity}

\noindent
For field transformations $\Phi'=\Phi+ \delta\Phi$ that do \emph{not
change} the path integral measure\footnote{Here only the anomaly free case is
considered. If the path integral measure is changed under the transformation
additional contributions must be taken into account.} ($\Dd\Phi=\Dd\Phi'$) Ward
identities naturally arise on the level of observables as
\begin{equation}
\vev{\mathcal{O}} = \ZZ^{-1}\!\!\int \Dd\Phi'\, \mathcal{O}[\Phi']\,
\ee^{-S[\Phi']} = \ZZ^{-1}\!\! \int \Dd\Phi\, (\mathcal{O}[\Phi]+\delta
\mathcal{O}[\Phi])(1-\delta S[\Phi])\, \ee^{-S[\Phi]} =
\vev{\mathcal{O}+\delta \mathcal{O}- \mathcal{O}\,\delta S},
\end{equation}
implying $\vev{\delta \mathcal{O}}=\vev{\mathcal{O}\,\delta S}$. If further the
action \emph{and} ground state are invariant under the transformation given by $\delta$ 
then $\vev{\delta \mathcal{O}}$ vanishes for every observable $\mathcal{O}$.

For unbroken supersymmetric theories Ward identities are used to test the
supersymmetry restoration in the continuum limit by analysing the continuum
limit of $\vev{\delta \mathcal{O}}$ for a given observables $\mathcal{O}$. If
supersymmetry is broken then Ward identities will \emph{not} be fulfilled in the continuum
limit,  $\vev{\delta \mathcal{O}}\neq 0$. On the lattice supersymmetry will be
further broken explicitly by a finite lattice spacing and by finite
temperature.

\begin{figure}\hfill
\includegraphics{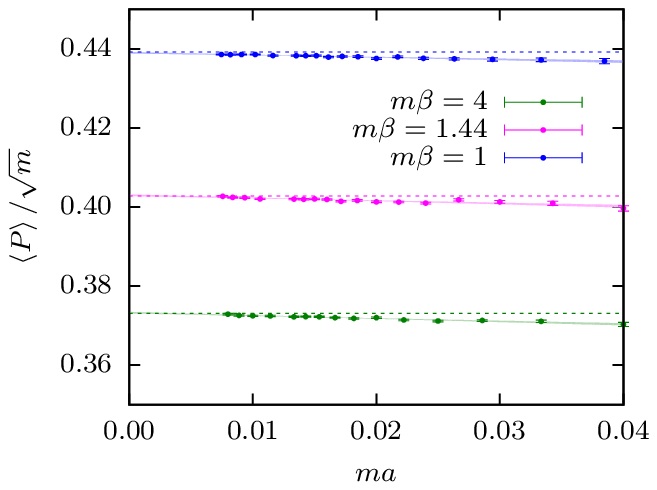}
\hfill
\includegraphics{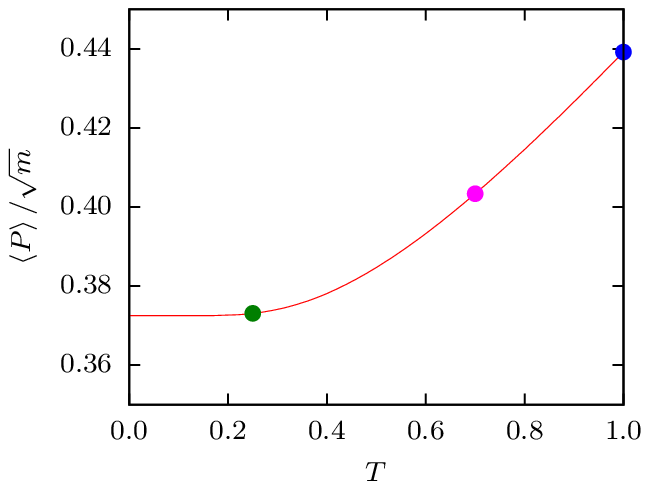}
\hfill\hfill\phantom.
\caption{\label{fig:qm:wardIdentityContinuum} Prepotential for coupling
$f=1$. Left panel: Continuum limit of the prepotential
$\vev{P}/\sqrt{m}$ which serves as simplest Ward identity for the
supersymmetric quantum mechanics. (The dashed lines denote the continuum values
while the shaded area gives the error bound of the linear extrapolation to
$a=0$.) Right panel: Temperature dependence of of $\vev{P}/\sqrt{m}$ computed
from the diagonalised Hamiltonian (data from the left panel is marked by points).}
\end{figure}

A simple Ward identity is provided by
\begin{equation}
\int \dd t\, \vev{\delta^{(1)}\bar\psi} = -\bar\varepsilon
\int \dd t\, \vev{\dot \phi+P(\phi)} = -\bar\varepsilon 
\vev{\int \dd t\, P(\phi)} = -\beta\bar \varepsilon\vev{P(\phi)},
\end{equation}
and a dimensionless identity is given by $\vev{P}/\sqrt{m}=0$ iff the
ground state is invariant under the supersymmetry. The impact of the explicit
supersymmetry breaking on the continuum result at vanishing temperature is
depicted in Fig.~\ref{fig:qm:wardIdentityContinuum} for coupling $f=1$. 
The finite $a$ effects are small for the considered lattice spacings and a linear
extrapolation to $a=0$ provides results in full agreement with the
continuum results (see Tab.~\ref{tab:sqm:wardContinuum}). Finite temperature
effects are negligible for $T<0.2$. The numerically exact value from the 
operator formalism is $\vev{P}/\sqrt{m}=0.37251$ and  corresponds to the 
non-supersymmetric ground states. This quantity is invariant under the $\bbZ_2$ 
symmetry $\Phi\to \Phi-1/f$
and is not sensitive to the specific ground state chosen at $T=0$.
\begin{table}
\centering
\begin{tabular}{c@{\hspace{5ex}}c@{\hspace{5ex}}c} \hline
$T^{-1}$ & exact & extrapolated \\ \hline
$1.00$ & $0.4392039$ & $0.43902(18)$ \\
$1.44$ & $0.4027741$ & $0.40298(20)$ \\
$4.00$ & $0.3730929$ & $0.37327(18)$ \\ \hline
\end{tabular}
\caption{\label{tab:sqm:wardContinuum} Linearly extrapolated lattice values of
the Ward identity $\vev{P}/\sqrt{m}$ in comparison to the exactly calculable ones
from the operator formalism for three different temperatures $T$ at coupling
$f=1$.}
\end{table}


\section{\boldmath$\mathcal{N}=1$ Wess-Zumino model}
\label{chap:n1wz}
\noindent
Supersymmetry as incorporated in the minimal
supersymmetric standard model \cite{Dimopoulos:1981zb} or extensions
thereof \cite{Ellwanger:2009dp} can only be a fundamental symmetry of nature
if it is spontaneously broken on experimentally accessible energy scales 
with a phase transition at a much higher energy. Guided
by this observation there is a need to study supersymmetry breaking phase
transitions with non-perturbative tools. To explore the possibilities provided by
lattice regularisations a minimal model is chosen, namely the
$\mathcal{N}=(1,1)$ Wess-Zumino model in two dimensions
\cite{Ferrara:1975nf}. It is obtained from the
$\mathcal{N}=(2,2)$ Wess-Zumino model, which provides a dimensionally reduced version
of the matter sector of the minimal supersymmetric standard model, by
constraining the fields of the $\mathcal{N}=(2,2)$ model to be real
\cite{Schiller:1986zx,Catterall:2001fr,Bergner:2007pu,Kastner:2007gz,Kastner:2008zc,Wozar:2008jb}.
This amounts to turning complex scalars into real ones and replacing Dirac fermions with Majorana
fermions, leading to a minimal field content with only one bosonic and fermionic
degree of freedom.\footnote{From a practical point of view the absence of gauge
fields has the advantage that derivatives can be applied in momentum space to speed up
Monte-Carlo simulations considerably.} For models with short range
interactions at least two dimensions are necessary to observe a phase transition. 
In particular supersymmetric quantum mechanics is not sufficient to
model a phase transition since the systems reside in one specific 
phase depending on the highest power of the superpotential (cf.~Sec.~\ref{sec:qm:opForm}). Since the seminal work by Witten
\cite{Witten:1982df} it is known that the index $\Tr (-1)^{N_\text{F}}$ can vanish
for specific choices of the prepotential
and supersymmetry may be broken spontaneously depending on the couplings of the
prepotential for the $\mathcal{N}=1$ Wess-Zumino model.

\begin{figure}
\centering
\includegraphics{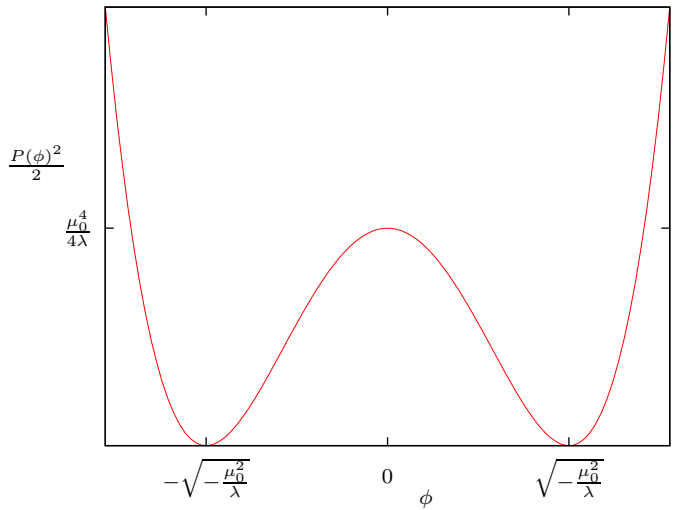}
\caption{\label{fig:classicalPotential} The classical bosonic potential
of the Wess-Zumino model given in Eq.~\eqref{eq:n1wz:classPotCont}.}
\end{figure}

The off-shell continuum formulation is given by the action
\begin{equation}
S = \int \dd^2 x\, \frac{1}{2} \bigl( (\partial_\mu \phi)^2 +\bar\psi (\slashed
\partial + P' (\phi)) \psi + 2 F P(\phi)-F^2 \bigr),
\end{equation}
where $F$ and $\phi$ denote real scalar fields and $\psi$ is a 
(real) Majorana spinor with two spinorial components. This formulation
is invariant under the supersymmetry transformations
\begin{equation}
\delta \phi = \bar\epsilon \psi,\quad
\delta \psi = (\slashed \partial\phi-F)\epsilon,\quad
\delta F = -\bar\epsilon \slashed \partial \psi.
\end{equation}
By eliminating the auxiliary field $F=P(\phi)$ one arrives at
the on-shell action
\begin{equation}
\label{eq:n1wz:contAction}
S = \int \dd^2 x\, \frac{1}{2} \left( (\partial_\mu \phi)^2 +\bar\psi (\slashed
\partial + P' (\phi)) \psi + P(\phi)^2 \right) .
\end{equation}
Here, the prepotential $P$ is chosen to be
\begin{equation}
\label{eq:n1wz:prepot}
P(\phi) = \frac{\mu_0^2}{\sqrt{2\lambda}}+\sqrt{\frac{\lambda}{2}}\phi^2,
\end{equation}
so that the classical potential (Fig.~\ref{fig:classicalPotential}) for the
scalar part corresponds to a $\phi^4$ theory with $\bbZ_2$ symmetry
($\phi\to-\phi$),
\begin{equation}
\label{eq:n1wz:classPotCont}
\frac{P(\phi)^2}{2} = \frac{\mu_0^2}{2} \phi^2 + \frac{\lambda}{4}\phi^4 +
\const,
\end{equation}
while the fermions acquire a Yukawa interaction. An analysis of the Witten index
\cite{Witten:1982df} reveals one bosonic and one fermionic ground state
that imply $\Tr (-1)^{N_\text{F}}=0$. This ground state structure then allows
for a spontaneous supersymmetry breaking. It is expected that for fixed
$\lambda>0$ and $\mu_0^2\ll 0$ the system cannot tunnel between the
two ground states so that supersymmetry is unbroken. On the other hand, for $\mu_0^2>0$
both ground state energies are lifted above zero and supersymmetry is broken. In consequence there will be a supersymmetry breaking phase transition for some $\mu_0^2<0$
and this section is devoted to the study of this transition.

There have already been several studies aiming at analysing and understanding
supersymmetry breaking in this model. By a strong coupling expansion
\cite{Bartels:1983wm} the occurrence of a supersymmetry breaking phase transition
was predicted. Calculations of the ground state energy with Monte-Carlo methods
\cite{Ranft:1983ag} confirm this expectation. Afterwards lower bounds on the
ground state energy have been analysed \cite{Beccaria:2004pa,Beccaria:2004ds} to
obtain a phase diagram of supersymmetry breaking by working in the Hamiltonian
formalism and making a numerical analysis with Green's function Monte-Carlo
methods. However, the obtained critical lattice couplings so far are only
unrenormalised couplings corresponding to one specific lattice spacing. Recently an analysis
based on exact renormalisation group methods \cite{Wetterich:1992yh} has been
performed \cite{Synatschke:2009nm} where no supersymmetry breaking is introduced
during the renormalisation group flow. Again, the phase diagram has been
obtained and supersymmetry breaking is found to coincide with a restoration of the
$\bbZ_2$ symmetry in a \emph{second order} phase transition. A supersymmetric
(massive) phase is found for weakly coupled systems as well as a phase of broken supersymmetry with 
\emph{goldstinos}, the massless Goldstone fermions \cite{Salam:1974zb} of
the broken supersymmetry\footnote{This breaking is \emph{not} forbidden by the
Mermin-Wagner theorem \cite{Mermin:1966fe} that only applies to \emph{bosonic}
symmetry generators.}, and bosons whose mass vanishes with growing
renormalisation group scale.\footnote{Also the corresponding critical exponents have been determined in \cite{Synatschke:2009nm}.} But still the critical coupling depends
on the chosen regulator which prevents a direct comparison of numerical values.

In general a naive discretisation of a lattice action has to face the problem of
broken supersymmetry for finite lattice spacing with the need to fine-tune
lattice couplings to reach a supersymmetric continuum limit. However, this model
has the advantage that the necessary counterterms have been analysed in lattice
perturbation theory \cite{Golterman:1988ta} and a lattice prescription is given
that ensures the supersymmetric continuum limit (perturbatively). Simulations of
the given discretisation (based on the Wilson derivative) have already
been performed \cite{Catterall:2003ae} and a tunnelling between the possible
ground states is found to coincide with the onset of supersymmetry breaking and
the appearance of a goldstino. Despite all these successes several open issues
remain: The
breaking of a $\bbZ_2$ symmetry (which is correlated with the restoration of
supersymmetry) has so far only been analysed with a $\bbZ_2$ breaking
action.\footnote{The Wilson term for the fermionic part of the action will break the $\bbZ_2$ symmetry
as analysed in \cite{Kastner:2008zc}.} Further the given critical coupling is
still regulator dependent and not directly comparable to other methods. For
that reason the aim of this section is the non-perturbative determination from
first principles of a renormalised critical coupling in the continuum limit.


\subsection{Quenched model}
\noindent
Although our primary focus lies on the supersymmetric model
it is useful to exemplify the definition of a renormalised critical coupling in
the setting of the quenched model where fermionic contributions are neglected.
This becomes especially important because a non-standard discretisation based on
the SLAC derivative is used.

With the prepotential of Eq.~\eqref{eq:n1wz:prepot} the quenched model is nothing
but the two dimensional $\phi^4$ model with action
\begin{equation}
\SB = \int \dd^2 x\, \frac{1}{2} \left( (\partial_\mu \phi)^2 + \mu_0^2
\phi^2 + \frac{\lambda}{2} \phi^4 \right).
\end{equation}
This model is (classically) invariant under a discrete $\bbZ_2$ symmetry
($\phi\to -\phi$) which can be broken dynamically in the full quantum theory
\cite{Chang:1976ek}. The symmetric phase is characterised by $\vev{\phi}=0$,
whereas in the broken phase (in the thermodynamic limit) $\vev{\phi}\neq 0$.

\begin{figure}
\centering
\includegraphics{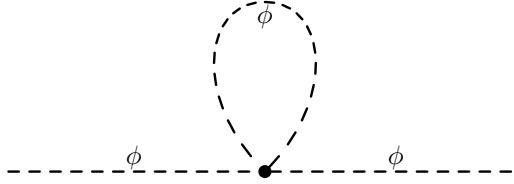}
\caption{\label{fig:leafDiagram} The only divergent Feynman diagram for the
bosonic $\phi^4$ model in the $\bbZ_2$ symmetric phase.}
\end{figure}%
In contrast to the full $\mathcal{N}=2$ Wess-Zumino model the $\phi^4$ model (as
well as the full $\mathcal{N}=1$ Wess-Zumino model) is not finite and there is
need for a renormalisation of couplings. In the 
$\bbZ_2$ symmetric phase the only divergence arises from the `leaf' diagram (see
Fig.~\ref{fig:leafDiagram}) and the model can be made finite with a mass renormalisation,
\begin{equation}
\SB = \int \dd^2 x\, \frac{1}{2} \left( (\partial_\mu \phi)^2 + \mu^2 \phi^2
+ \frac{\lambda}{2} \phi^4  - \delta \mu^2\, \phi^2\right).
\end{equation}
The (one-loop) relation between the inverse propagators is given in terms of
renormalised couplings by
\begin{equation}
G^{-1}(p) =  p^2 + \mu^2 + \Sigma(p^2),\quad
\Sigma(p^2) = 3 \lambda A_{\mu^2} - \delta \mu^2,\quad
A_{\mu^2} = \int \frac{\dd^2 p}{(2\pi)^2}\frac{1}{p^2+\mu^2}
\end{equation}
and the only (logarithmically) divergent contribution is $A_{\mu^2}$.
Hence all ultraviolet divergences can be removed by a 
renormalisation of the mass term\footnote{Equivalently this renormalisation can
be obtained by a normal ordering of the interaction part in the symmetric phase,
\begin{equation*}
\SB = \int \dd^2 x\, \frac{1}{2} \left( (\partial_\mu \phi)^2 + \mu^2 \phi^2
+ \frac{\lambda}{2} \colon\!\phi^4\!\colon\!_{\mu} \right),
\end{equation*}
where $\colon.\,\colon\!_{\mu}$ denotes normal ordering with respect to the
mass $\mu$.},
\begin{equation}
\delta\mu^2 = 3\lambda A_{\mu^2} \quad \Leftrightarrow\quad \mu_0^2 =
\mu^2-3\lambda A_{\mu^2}.
\end{equation}
This renormalisation prescription will also be sufficient in the $\bbZ_2$ 
broken phase \cite{Chang:1976ek}. We follow \cite{Loinaz:1997az} and use
the dimensionless renormalised coupling $f=\lambda/\mu^2$ that
distinguishes between the symmetric and broken phase. 
By contrast, (naively) possible definitions of a
renormalised coupling that are based on the pole mass or the propagator at
vanishing momentum are \emph{not sufficient} to distinguish between both phases.
Note that this renormalisation prescription can only be applied in a given
scheme and continuum results will then follow by removing the
ultraviolet regulator.\footnote{There is no need to introduce an infrared
regulator.}


\subsubsection{The $\bbZ_2$ phase transition on the lattice}

\noindent
The computation of the critical coupling in the bosonic $\phi^4$ model has a
long history where several methods and approximations (e.g.\ based on the
Gaussian effective potential or light-cone quantisation) have contributed.
For a recent overview containing the most precise Monte-Carlo results see
\cite{Schaich:2009jk}. Since these accurate results serve as reference values 
in the present work we briefly recall the most important points.
The lattice regularisation in
\cite{Loinaz:1997az,Schaich:2009jk} is based on the ``canonical'' discretisation utilising the forward derivative
and a renormalised critical coupling is computed with high precision. The canonical
model is given by
\begin{equation}
\SB = \sum_x \frac{1}{2}\left(\sum_\nu (\phi_{x+\hat\nu}-\phi_x)^2 + \hat\mu_0^2
\phi_x^2 + \frac{\hat\lambda}{2} \phi_x^4 \right)
\end{equation}
with dimensionless lattice parameters $\hat\lambda = \lambda a^2$ and
$\hat\mu_0^2 = \mu_0^2 a^2$.
Since $\lambda$ acquires \emph{no renormalisation} it is used to set the scale.
Equivalently $\hat\lambda$ determines the lattice spacing with $\hat\lambda\to
0$ in the continuum limit. The (dimensionless) renormalised coupling is again
given by $\hat f = \hat\lambda/\hat\mu^2$ and the corresponding
$\hat\mu_0^2$ can be computed via $\hat\mu_0^2=\hat\mu^2-3\hat\lambda
A_{\hat\mu^2}$, where $A_{\hat\mu^2}$ that enters the normal ordering is given in the infinite
volume limit by the lattice propagator for the forward derivative,
\begin{equation}
A_{\hat\mu^2} = \lim_{n\to \infty} n^{-2}\sum_{k_1=1}^{n}\sum_{k_1=1}^{n}
\frac{1}{\hat\mu^2 + 4\sin^2(\pi k_1/n) + 4\sin^2(\pi k_2/n)}.
\end{equation}
Because $A_{\hat\mu^2}$ diverges only logarithmically for $a\to 0$ it follows
that $\hat\mu^2,\hat\mu_0^2\to 0$ in the continuum limit at fixed $\hat f$. One
can show that at every fixed $\hat\lambda$ a second order phase transition appears.
However, the continuum physics at these phase transitions corresponds to an
\emph{infinite} continuum coupling $\lambda$ and is therefore
only \emph{indirectly relevant} for the continuum $\phi^4$ model under
consideration.

\begin{figure}
\centering
\includegraphics{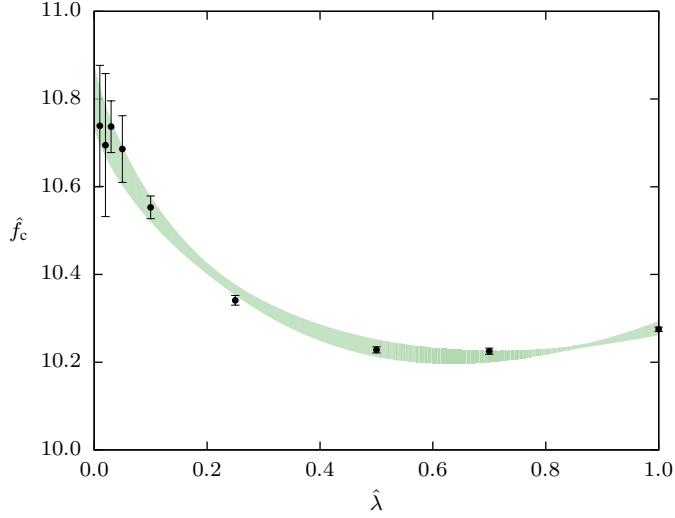}
\caption{\label{fig:criticalCouplingsQuenched} Extrapolation of the critical
couplings of the bosonic $\phi^4$ model to the continuum limit. The shaded
region indicates error bands for the extrapolation using the functional form
\eqref{eq:quenchedContinuumExtrapolation}. Data points are taken from
\cite{Schaich:2009jk} with the canonical discretisation while confidence bands
are computed here.}
\end{figure}%
At finite lattice spacing (given by fixed $\hat\lambda$) there will be a
$\bbZ_2$ breaking phase transition and a critical $\hat\mu^2_\text{c}$ can be
extracted. The renormalised critical coupling in the continuum limit is then
determined via
\begin{equation}
\label{eq:n1:criticalCouplingDef}
  f_\text{c} = \left[\frac{\lambda}{\mu^2}\right]_\text{crit} =
  \lim_{\hat\lambda\to 0} \hat f_\text{c}\quad \text{with}\quad \hat f_\text{c}=
  \frac{\hat\lambda}{\hat\mu^2_\text{c}}.
\end{equation}
The phase transition itself for finite $\hat\lambda$ can be determined from
the Binder cumulant
\begin{equation}
U =
1-\frac{\bigl\langle\tilde\phi^4\bigr\rangle}{3\bigl\langle\tilde\phi^2\bigr\rangle^2}
\quad\text{with}\quad \tilde\phi=N^{-1}\sum_x\phi_x,
\end{equation}
which becomes \emph{independent} of the
lattice volume \emph{at} the second order phase transition point \cite{Binder:1981sa}.\footnote{Strictly speaking there is
still a slight volume dependence such that the large volume extrapolation of the
intersection points for different lattice volumes corresponds to the phase
transition.} The critical coupling has been determined from lattices up to a size of
$1200^2$ and was found to be affected by linear and logarithmic corrections 
in the lattice spacing. An extrapolation based on the published values in
\cite{Schaich:2009jk} for $\hat\lambda\in[0.01,1]$ to the continuum using a
functional form
\begin{equation}\label{eq:quenchedContinuumExtrapolation}
\hat f_\text{c}(\hat\lambda) \approx f_\text{c} + a \hat\lambda + b\hat\lambda
\ln \hat\lambda
\end{equation}
reveals a renormalised critical coupling in the continuum of 
$f_\text{c}= 10.81(7)$ (see Fig.~\ref{fig:criticalCouplingsQuenched}).


\subsubsection{Regulator independence of the renormalised critical coupling}
\noindent
The results of Sec.~\ref{chap:qm} and results in
\cite{Bergner:2007pu,Kastner:2007gz,Kastner:2008zc,Wozar:2008jb,Bergner:2009vg}
imply that a discretisation based on the SLAC derivative gives results close to the continuum limit
and will not break the $\bbZ_2$ symmetry of the full supersymmetric model. For that
reason the SLAC derivative will also be used to simulate the $\mathcal{N}=1$
Wess-Zumino model. In contrast to other low dimensional scalar supersymmetric
models it is now necessary to cope with a logarithmic mass renormalisation and the renormalised
lattice coupling depends on the chosen renormalisation procedure.

To justify the applicability of the SLAC derivative also for the present case
simulations based on the lattice action
\begin{equation}
S = \sum_x \frac{1}{2}\left(\sum_\nu (\partial^\text{SLAC}_\nu\phi)_x^2 +
\hat\mu_0^2 \phi_x^2 + \frac{\hat\lambda}{2} \phi_x^4\right)
\end{equation}
have been performed where the mass parameter is still given by
$\hat\mu_0^2=\hat\mu^2-3\hat\lambda A_{\hat\mu^2}$. Only now, $A_{\hat\mu^2}$ 
is determined from the propagator based on the SLAC derivative,
\begin{equation}
\label{eq:n1wz:renormalisationSlac}
A_{\hat\mu^2} = \lim_{n\to \infty}
(2n)^{-2}\sum_{k_1=-n+1}^{n}\sum_{k_2=-n+1}^{n} \frac{1}{\hat\mu^2 + (\pi k_1/n)^2 + (\pi k_2/n)^2}.
\end{equation}

\begin{figure}\hfill
\includegraphics{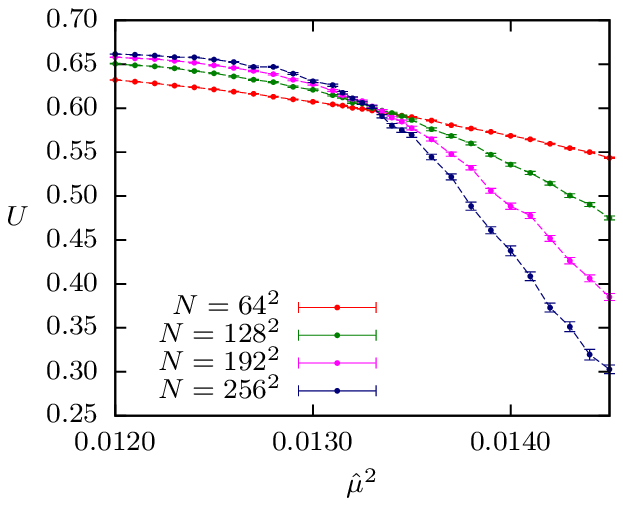}
\hfill
\includegraphics{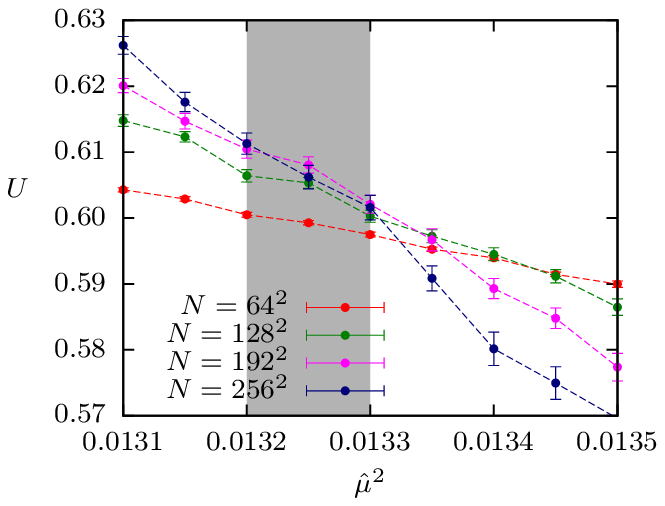}
\hfill\hfill\phantom.
\caption{\label{fig:binderQuenched15} Binder cumulants at
$\hat\lambda=0.15$ for different lattice sizes. The coarse view is given in
the left panel while a close-up view of the critical region is shown in the
right panel. From the intersection point for the largest lattices a critical
coupling $\hat f_\text{c}=\frac{0.15}{0.01325(5)}=11.321(43)$ is extracted. The
shaded region gives the error band for the infinite volume extrapolation.}
\end{figure}
\begin{figure}
\centering
\includegraphics{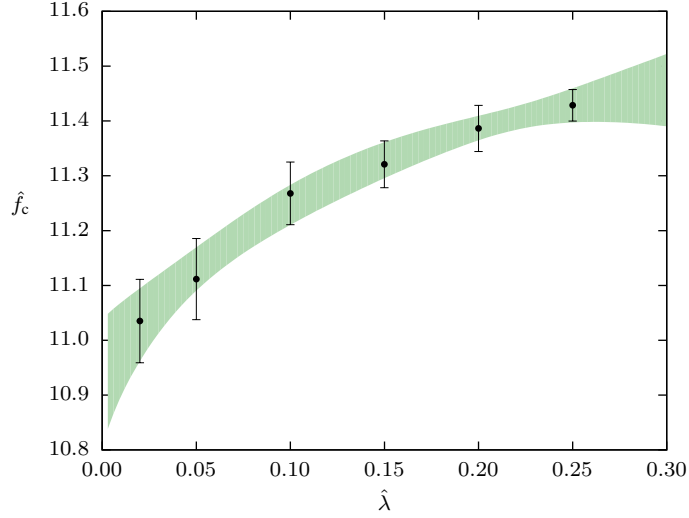}
\caption{\label{fig:criticalCouplingsQuenchedSl} Extrapolation of the critical
couplings of the bosonic $\phi^4$ model based on the SLAC derivative to the
continuum limit according to the functional form
\eqref{eq:quenchedContinuumExtrapolation}.}
\end{figure}

%
\begin{table}
\centering
\begin{tabular}{c@{\hspace{5ex}}c} \hline
$\hat\lambda$ & $\hat f_\text{c}$ \\ \hline
$0.02$ & $11.035(76)$ \\
$0.05$ & $11.112(74)$ \\
$0.10$ & $11.268(57)$ \\
$0.15$ & $11.321(43)$ \\
$0.20$ & $11.386(42)$ \\
$0.25$ & $11.429(29)$ \\ \hline
\end{tabular}
\caption{\label{tab:n1:quenchedCritical}Renormalised critical couplings for the
$\phi^4$ model as determined from lattice sizes up to $256^2$ with the SLAC
derivative.}
\end{table}
Similar to the case of the naive discretisation the crossing of the Binder
cumulant $U$ for different lattice volumes at fixed $\hat\lambda$ determines the critical
$\hat\mu^2$ and therefore the critical coupling $\hat f_\text{c}$. This
procedure is exemplified for $\hat\lambda=0.15$ in Fig.~\ref{fig:binderQuenched15} where
lattice sizes up to $256^2$ were used. As an outcome of these calculations
critical couplings have been determined for six different $\hat\lambda$ (see
Tab.~\ref{tab:n1:quenchedCritical}). Again, an extrapolation to the continuum
limit $\hat\lambda\to 0$ has been performed according to
Eq.~\eqref{eq:quenchedContinuumExtrapolation} (see Fig.~\ref{fig:criticalCouplingsQuenchedSl}) and gives the
renormalised critical coupling in the continuum of $f_\text{c} = 10.92(13)$ which is in complete
agreement with the reference value $f_\text{c}=10.81(7)$ of
\cite{Schaich:2009jk}. This proves that the lattice $\phi^4$ model 
with non-local SLAC derivative possesses the correct continuum limit,
as expected from the analytic studies in \cite{Bergner:2007pu,bergner:2009phd}. 
The definition of a renormalised critical continuum coupling is \emph{independent}
of the chosen lattice regulator.


\subsection{Full dynamical model}
\label{sec:n1:full}
\noindent
By inclusion of dynamical fermions the model
is now given in the continuum by the action \eqref{eq:n1wz:contAction}.
Irrespective of the chosen prepotential $P(\phi)$ the action is invariant
under one supersymmetry. The dynamical breaking of this
supersymmetry for the prepotential defined in Eq.~\eqref{eq:n1wz:prepot} will be analysed in the
following.


\subsubsection{Renormalised lattice parameters}
\noindent
Using a discretisation
based on the SLAC derivative the lattice action is given by a direct
discretisation of the corresponding continuum action,
\begin{equation}
S = \sum_x \frac{1}{2}\Bigl(\sum_\nu (\partial^\text{SLAC}_\nu\phi)_x^2 +
\hat\mu_0^2 \phi_x^2 + \frac{\hat\lambda}{2} \phi_x^4 +\sum_y
\psi^\trnsp_x\underbrace{C(\slashed\partial^\text{SLAC}_{xy}+\sqrt{2\hat\lambda}\phi_x\delta_{xy})}_{=M[\phi]}
\psi_y \Bigr).
\end{equation}
In a Majorana representation with $\gamma$ matrices and charge conjugation matrix
\begin{equation}
\gamma_0 = \begin{pmatrix}
           1 & 0 \\ 0 & -1
           \end{pmatrix}, \quad
\gamma_1 = \begin{pmatrix}
           0 & -1 \\ -1 & 0
           \end{pmatrix}, \quad
C = \begin{pmatrix}
           0 & -1 \\ 1 & 0
           \end{pmatrix}
\end{equation}
is the fermion matrix $M$ real and antisymmetric.

In \cite{Golterman:1988ta} a lattice model of the $\mathcal{N}=1$ Wess-Zumino
model built upon the symmetric derivative has been analysed. To avoid
the species doubling problem a Wilson mass term has been added to the
prepotential. Similar discretisations were used for the $\mathcal{N}=2$ 
Wess-Zumino model in
\cite{Catterall:2001fr,Bergner:2007pu,Kastner:2007gz,Kastner:2008zc,Wozar:2008jb}. With
properly renormalised coupling parameter (as described below) is was shown that the
supersymmetric continuum limit is reached. Although this restoration was only
analysed for a discretisation based on Wilson fermions the whole
line of argument is directly applicable for the SLAC derivative. This follows
from the fact that the SLAC derivative for two dimensional models with
Yukawa interaction needs no non-local or non-covariant counterterms to construct
a local continuum limit and the lattice degree of divergence of Feynman
diagrams coincides with the divergence of the corresponding continuum
diagrams \cite{Bergner:2007pu,Kadoh:2009sp,Bergner:2009vg}.

\begin{figure}
\centering
\includegraphics{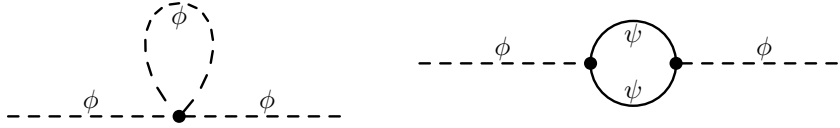}
\caption{\label{fig:n1:fullDiagrams} The divergent Feynman diagrams for the full
$\mathcal{N}=1$ Wess-Zumino model in the $\bbZ_2$ symmetric phase.}
\end{figure}

The analysis of divergent diagrams in \cite{Golterman:1988ta} starts from the
$\bbZ_2$ broken phase and shows that a logarithmic renormalisation of the bare
mass parameter is necessary to cancel divergent contributions,
\begin{equation}
\hat\mu_0^2=\hat\mu^2+\frac{\hat\lambda}{4\pi}\left(\ln \hat\mu^2+c\right),
\end{equation}
where $c$ may be any constant to fix the renormalisation scale. In analogy to
the quenched model this constant is now fixed to obtain
\begin{equation}
\label{eq:n1:bareCouplingFull}
\hat\mu_0^2=\hat\mu^2-1 \hat\lambda
  A_{\hat\mu^2}
\end{equation}
with $A_{\hat\mu^2}$ defined for the SLAC derivative in
Eq.~\eqref{eq:n1wz:renormalisationSlac}. In contrast to the quenched model only
a factor `$1$' in front of the divergent part is needed that arises from a
partial cancellation of the Feynman diagrams shown in
Fig.~\ref{fig:n1:fullDiagrams}. Compared to the $\mathcal{N}=2$ Wess-Zumino model 
there is \emph{no complete cancellation} and
a divergence remains. Again the given renormalisation procedure amounts to a
normal ordering of interaction terms with mass parameter $\hat\mu$ in the
$\bbZ_2$ symmetric phase. Eventually a renormalised coupling is defined in the
continuum limit similarly as in the bosonic case according to
Eq.~\eqref{eq:n1:criticalCouplingDef}, but now with $\hat\mu^2_0$ given through
Eq.~\eqref{eq:n1:bareCouplingFull}.

\subsubsection{The Pfaffian}
\label{ssec:n1:pfaffian}
\noindent
The field content of the $\mathcal{N} = 1$ Wess-Zumino model includes Majorana
fermions for which the fermionic path integral yields a Pfaffian,
\begin{equation}
\ZZ = \int \DD\phi\,\DD \psi\, \ee^{-\SB[\psi]-\psi^\trnsp M[\phi] \psi} = \int
\DD\phi\, \Pf M[\phi]\, \ee^{-\SB[\phi]}.
\end{equation}
where $M[\phi]$ is an antisymmetric matrix.\footnote{In general, as it is the
case e.g.\ for supersymmetric Yang-Mills theories \cite{Montvay:2001ry}, the
matrix $M$ is not necessarily real. However, it is still antisymmetric
but does not need to be anti-Hermitian.} The Pfaffian
is (up to a sign) the square root of the determinant, $(\Pf M )^2 = \det M$, and follows from Grassmann integration,
\begin{equation}
\Pf M = \frac{1}{2^N N!}\sum_{\sigma\in S_{2N}} \sign(\sigma) \prod_{i=1}^{N}
M_{\sigma_{2i-1},\sigma_{2i}}.
\end{equation}
In practice the Pfaffian is computed as described in \cite{Jia:2006,Rubow:2011dq} 
with complexity $\ord(N^3)$. Similar to the case of Dirac fermions in
Sec.~\ref{ssec:qm:sgnDet} this Pfaffian may have a fluctuating sign and Monte-Carlo
simulations are carried out with the effective action
\begin{equation}
\Seff = \SB - \ln \abs{\Pf M} = \SB - \frac{1}{2}\ln \det M\quad \Rightarrow
\quad \ZZ = \int
\DD\phi\, \ee^{-\Seff[\phi]}.
\end{equation}
Nevertheless, the sign of the Pfaffian must be taken into account by
reweighting of measurements.

With the chosen representation of the Clifford algebra the fermion matrix is
given by 
\begin{equation}
M = \begin{pmatrix}
    \partial_1^\text{SLAC} & \partial_0^\text{SLAC}-P' \\
    \partial_0^\text{SLAC}+P' & -\partial_1^\text{SLAC}
    \end{pmatrix}.
\end{equation}
By using general identities for the Pfaffian of an antisymmetric
matrix $A\in \mathbb{R}^{2n\times 2n}$ and general matrix $B\in
\mathbb{R}^{2n\times 2n}$,
\begin{equation}
\label{eq:n1:paffianRelations}
\Pf(A) = (-)^n \Pf(A^\trnsp),\quad
\Pf(BAB^\trnsp) = \det(B)\Pf(A),
\end{equation}
transformation properties under the $\bbZ_2$ symmetry
$\phi\to-\phi$ of the bosonic potential $P^2/2$ can be derived.
For any lattice-derivative with antisymmetric matrix representation
(e.g.\ for the SLAC derivative) on a lattice with $N$ points
\begin{equation}
\begin{aligned}
\Pf \begin{pmatrix}
    \partial_1 & \partial_0-P' \\
    \partial_0+P' & -\partial_1
    \end{pmatrix}& \overset{\text{transpose}}=
(-)^N\Pf\begin{pmatrix}
    -\partial_1 & -\partial_0+P' \\
    -\partial_0-P' & \partial_1
    \end{pmatrix} \\
&\overset{\begin{pmatrix}
          0 & -1\\ 1 & 0
          \end{pmatrix}}=(-)^N\Pf\begin{pmatrix}
    \partial_1 & \partial_0+P' \\
    \partial_0-P' & -\partial_1
    \end{pmatrix}
\end{aligned}
\end{equation}
holds true. Under the $\bbZ_2$ transformation $P' \to -P'$ the Pfaffian is
\emph{invariant} on even lattice volumes and hence it defines a 
discrete symmetry of the full model. On the other hand, for an odd number of 
lattice sites the Pfaffian changes
sign under $P' \to -P'$. The seemingly inconsistent behaviour where symmetry
properties depend on the number of lattice points can be resolved by
a closer look at the SLAC derivative. For this derivative the
number of lattice points is directly related to the boundary conditions of the
fields. By imposing the natural condition that the spectrum of the lattice derivative
operator lies symmetric around the real axis in momentum space an even number
of lattice points \emph{must} be used for antiperiodic boundary conditions and
an odd number for periodic ones.\footnote{Nevertheless, the \emph{squared} SLAC
derivative can be defined unambiguously for periodic fields with an even
number of lattice points.} In consequence the Pfaffian changes sign
under $P'\to -P'$ for periodic boundary conditions. Then the contribution
of every field configuration with positive Pfaffian is cancelled in the path integral 
by the contribution of the transformed configuration with negative Pfaffian 
and the same bosonic action. This implies a vanishing Witten index, which 
is nothing but the path integral with periodic fermionic boundary conditions. 
For antiperiodic fermions in the
temporal direction the Pfaffian keeps its sign under the $\bbZ_2$ symmetry in
accordance with the positive definite partition function for the thermal ensemble.

For Wilson fermions the situation is different. In that 
case the derivative has still an antisymmetric
matrix representation but the contribution of the Wilson term leads to
\begin{equation}
\begin{aligned}
\Pf \begin{pmatrix}
    \partial_1 & \partial_0-P'-\frac{r}{2}\Delta \\
    \partial_0+P'+\frac{r}{2}\Delta & -\partial_1
    \end{pmatrix}& \overset{\text{transp.}}=
(-)^N \Pf\begin{pmatrix}
    -\partial_1 & -\partial_0+P'+\frac{r}{2}\Delta \\
    -\partial_0-P'-\frac{r}{2}\Delta & \partial_1
    \end{pmatrix} \\
&\overset{\begin{pmatrix}
          0 & -1\\ 1 & 0
          \end{pmatrix}}=(-)^N\Pf\begin{pmatrix}
    \partial_1 & \partial_0+P'+\frac{r}{2}\Delta \\
    \partial_0-P'-\frac{r}{2}\Delta & -\partial_1
    \end{pmatrix}.
\end{aligned}
\end{equation}
Changing the sign of $P'$ can in general only preserve the modulus 
of the Pfaffian if the Wilson parameter $r$ changes its sign, too. Since the
lattice theory is defined with fixed $r$ the $\bbZ_2$ symmetry will
be broken by the Wilson term, similar to the $\mathcal{N}=2$ Wess-Zumino
model \cite{Kastner:2008zc}. For that reason the discretisation based on the
SLAC derivative extends the works \cite{Golterman:1988ta,Catterall:2001wx,Catterall:2003ae} by
implementing the $\bbZ_2$ symmetry of the continuum model explicitly on the
lattice.


\subsubsection{Symmetries, boundary conditions, and ground states}
\label{ssec:n1:groundStates}
\noindent
The static properties under the $\bbZ_2$ symmetry that depend on the boundary
conditions for the dynamical fermions can be directly related to the ground state
structure of the full model. In the case of broken $\bbZ_2$ symmetry 
one ground state is correlated to positive $\tilde\phi$
while the other one is correlated to
negative $\tilde\phi$. The analysis of $\sign \Pf M$ reveals that in exactly one of these states the sign flips when one changes the
boundary conditions. A change of the fermionic boundary conditions 
amounts to an insertion of $(-1)^{N_\text{F}}$ into the path integral
which exactly implies the existence of one bosonic and one fermionic ground state.

\begin{figure}\hfill
\includegraphics{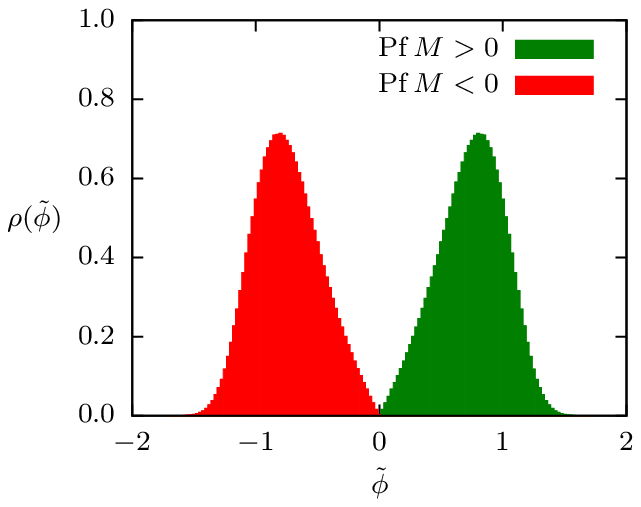}
\hfill
\includegraphics{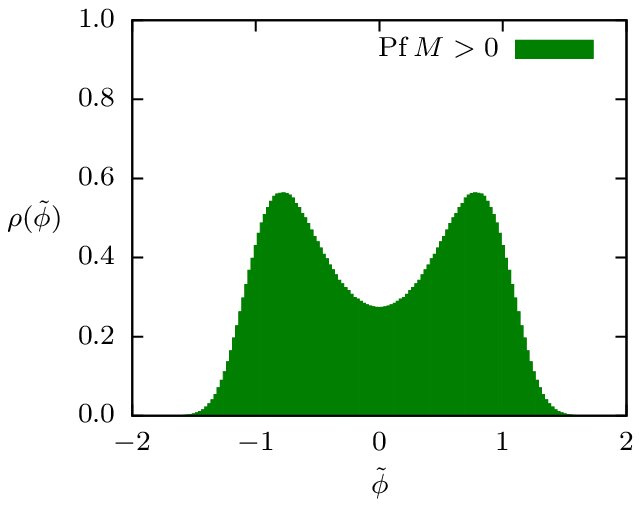}
\hfill\hfill\phantom.
\caption{\label{fig:histogramsBC} Probability density of the volume averaged
bosonic field for periodic boundary conditions on a $9\times 9$ lattice (left panel) and
thermal boundary conditions on an $8\times 9$ lattice (right panel) at
coupling $\hat f=100$ and $\hat\lambda=0.1$. The histograms are computed (with a
statistics of $6\cdot 10^6$ configurations) separately for 
fixed $\sign \Pf M$.}
\end{figure}%
These relations have been checked in the
dynamical ensemble with (small scale) lattice simulations with periodic and
antiperiodic temporal boundary conditions for the fermions at couplings
$\hat\lambda=0.1$ and $\hat f=100$ (see Fig.~\ref{fig:histogramsBC}). The
different boundary conditions at finite temperature imply that the functional forms of
both histograms need not necessarily coincide. However, all
configurations with $\tilde\phi>0$ keep $\sign \Pf M$ under a change of
fermionic boundary conditions, such that the bosonic ground state is located at
$\tilde\phi>0$, while the fermionic has a support  $\tilde\phi<0$.
These relations have been checked at further couplings $\hat f\in[10,100]$ and  
$\sign \tilde\phi\cdot \sign \Pf M>0$ is found on \emph{every} configuration.
Similar to the broken supersymmetric quantum mechanics in Sec.~\ref{chap:qm}
it is necessary to use thermal boundary conditions to have a positive measure.\footnote{The
Pfaffian is found to be strictly positive for finite temperature
simulations with several couplings on lattices up to $20\times 21$. However, for
even larger lattices the calculation of the Pfaffian is getting extremely time
consuming and a representative statistics has not been generated.}

This choice of thermal boundary conditions has further implications for the
analysis of supersymmetry breaking. Apart from the \emph{explicit} supersymmetry
breaking introduced in the lattice theory by the finite lattice spacing and
finite volume there is a further explicit breaking introduced by the finite
temperature. For that reason an analysis of the \emph{spontaneous}
supersymmetry breaking in the continuum theory will involve the limit of
infinite volume (``thermodynamic limit''), vanishing temperature, and vanishing
lattice spacing of the lattice theory.


\subsubsection{$\bbZ_2$ breaking}
\label{ssec:n1:z2breaking}
\begin{figure}\hfill
\includegraphics{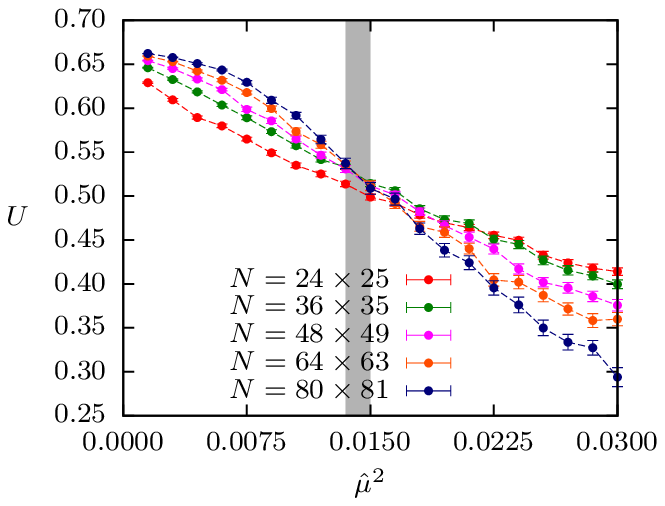}
\hfill
\includegraphics{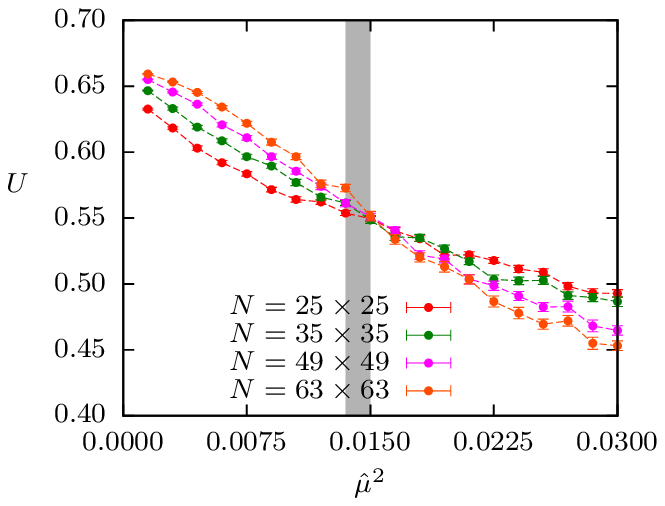}
\hfill\hfill\phantom.
\caption{\label{fig:binderSusy03}Binder cumulants for thermal (left panel) and
supersymmetry preserving (right panel) boundary conditions determined at fixed
$\hat\lambda=0.3$ and varying $\hat\mu^2$ (with a statistic of $10^4$
configurations). The shaded area denotes the error bounds of the critical
$\hat\mu^2$.}
\end{figure}

\noindent
The numerical survey of the phase diagram starts in analogy to the quenched
case with the determination of the phase transition associated to the $\bbZ_2$
breaking. The binder cumulant $U$ is thus computed for different volumes
at fixed lattice spacing with thermal boundary conditions. However,
a comparison to results obtained with periodic boundary conditions (without
reweighting) is enlightening. For $\hat\lambda=0.3$ the
intersection point of the Binder cumulants is independent of the chosen
boundary conditions (see Fig.~\ref{fig:binderSusy03}). This behaviour is
explained by the ground state structure at infinite lattice volume. In the
$\bbZ_2$ broken phase (for small $\hat\mu^2$) the system resides in only one
ground state with fixed $\sign \Pf M$. In that case expectation values are 
insensitive to the boundary conditions for the fermions and periodic boundary
may be imposed. For that reason it is quite safe to approach
the phase transition from the $\bbZ_2$ broken region and extract a critical
coupling from a crossing of Binder cumulants \emph{at the edge} of the
$\bbZ_2$ broken phase. But in the $\bbZ_2$ symmetric phase \emph{reweighted} 
expectation values are undefined for periodic boundary conditions  
and un-reweighted values may only be used with caution.

\begin{figure}\hfill
\includegraphics{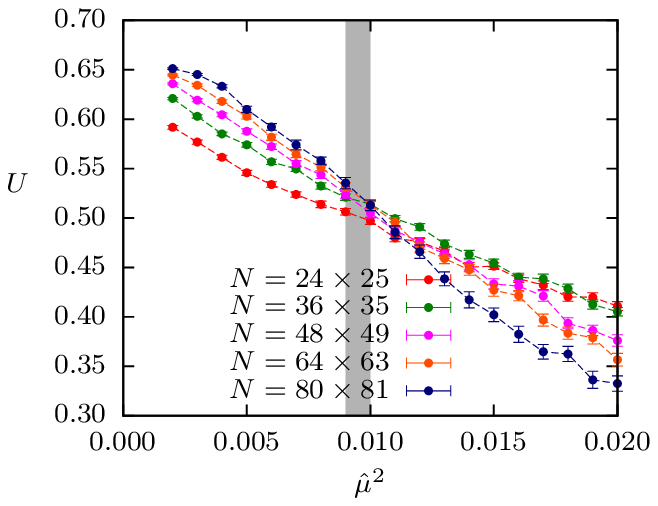}
\hfill
\includegraphics{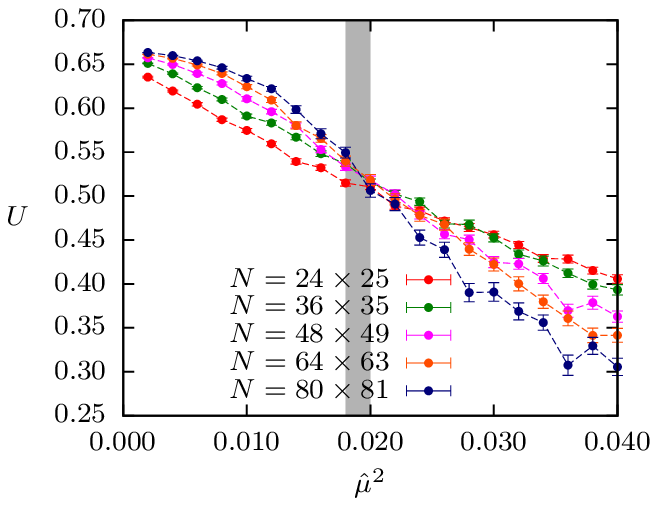}
\hfill\hfill\phantom.
\caption{\label{fig:binderSusy024}Binder cumulants for thermal
boundary conditions determined at fixed $\hat\lambda=0.2$ (left panel) and
$\hat\lambda=0.4$ (right panel) with a critical $\hat\mu^2$ indicated by the
shaded area.}
\end{figure}

The phase transition has also been determined for two further lattice spacings
with thermal boundary conditions (see Fig.~\ref{fig:binderSusy024}) and the
results for every $\hat\lambda$ are in full agreement with a critical coupling
of $f_\text{c}=21.1(1.1)$. At this point the numerical precision is
just not sufficient to resolve any running of the critical coupling with varying
lattice spacing, mainly because of the accessible lattice sizes. Hence the
determined critical couplings are used as continuum critical coupling
for the $\bbZ_2$ breaking with a broken phase for $f>f_\text{c}$ and a symmetric
phase for $f<f_\text{c}$.


\subsubsection{Supersymmetry breaking}
 
\noindent
As found in Sec.~\ref{ssec:n1:groundStates} the $\mathcal{N}=1$ Wess-Zumino
model with the chosen prepotential possesses one bosonic and one fermionic
ground state which are related through the $\bbZ_2$ symmetry. In the 
$\bbZ_2$ broken phase one definite ground state is selected and this
ground state will be the supersymmetric one because its partner state 
is not present in the physical spectrum at infinite volume. On the 
other hand, in the $\bbZ_2$ symmetric phase there may exist a 
supersymmetric ground state -- in contrast to supersymmetric 
quantum mechanics does the $\bbZ_2$ symmetry \emph{not necessarily exclude} a supersymmetric ground state. Thus it is necessary to study supersymmetry breaking 
on its own.

The direct way to study supersymmetry breaking is given by Ward
identities that are related to the supersymmetry transformation. If there is
one broken Ward identity then supersymmetry is broken. The simplest Ward
identity which is inherently related to the ground state energy is constructed from the
transformation of the fermionic field,
\begin{multline}
\label{eq:n1:prepot}
-\vev{V^{-1}\int\dd^2 x\, \delta\psi} = \epsilon \vev{V^{-1}\int \dd^2 x \,P} =
0 \quad \Leftrightarrow \quad \vev{\mathcal{P}} = 0\\
\text{with}\quad
\vev{\mathcal{P}} = \langle\hat{\mathcal{P}}\rangle/\sqrt{\hat\lambda},\quad
\hat{\mathcal{P}} = N^{-1}\sum_x\biggl(
\frac{\hat \mu_0^2}{\sqrt{2\hat
\lambda}}+\sqrt{\frac{\hat \lambda}{2}}\phi^2\biggr).
\end{multline}
It follows that the dimensionless prepotential $\mathcal{P}$ serves as an indicator
for supersymmetry breaking. However, supersymmetry is 
explicitly broken by the finite lattice spacing, finite
temperature and finite volume. Therefore dynamical supersymmetry breaking
should be examined in the limit of infinite lattice volume $N\to \infty$, in
combination with the continuum limit $a\to 0$.

\begin{figure}\hfill
\includegraphics{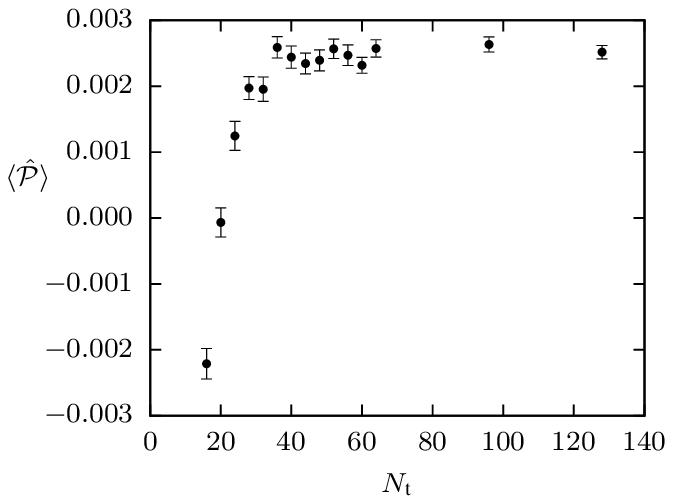}
\hfill
\includegraphics{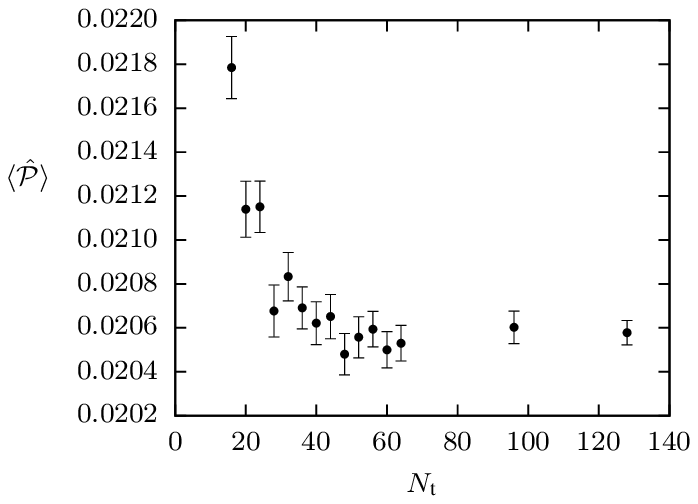}
\hfill\hfill\phantom.
\caption{\label{fig:n1:prepotVanishingT}Prepotential
$\langle\hat{\mathcal{P}}\rangle$ at fixed $\hat\lambda=0.1$ and $\Ns=35$ for
the $\bbZ_2$ broken phase with $\hat f=100$ (left panel) and the $\bbZ_2$
symmetric phase with $f=10$ (right panel).}
\end{figure}%
The first limit to be studied is the limit of vanishing temperature at fixed
lattice spacing ($\hat\lambda=0.1$) and fixed spatial volume ($\Ns=35$).
Simulations have been performed at $\hat f=100$ ($\bbZ_2$ unbroken) and
$\hat f=10$ ($\bbZ_2$ broken) and the results plotted in
Fig.~\ref{fig:n1:prepotVanishingT} indicate that for $\Nt\gtrsim \Ns$ the finite
temperature corrections become negligible in both phases. Hence in the
following nearly quadratic lattices are chosen with $\Nt=\Ns\pm
1$.\footnote{The choice of the sign in $\Ns\pm 1$ depends on the efficiency
of the involved Fourier transformations.} But already at this point the
prepotential is one order of magnitude larger in the $\bbZ_2$ symmetric phase than in the $\bbZ_2$ broken phase.

\begin{figure}\hfill
\includegraphics{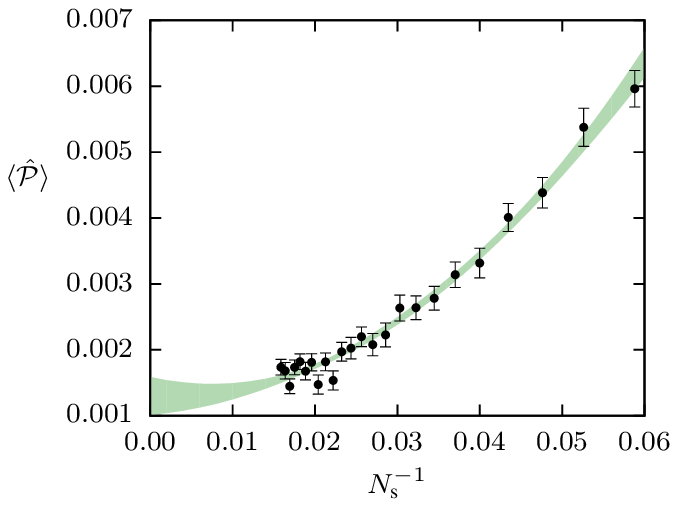}
\hfill
\includegraphics{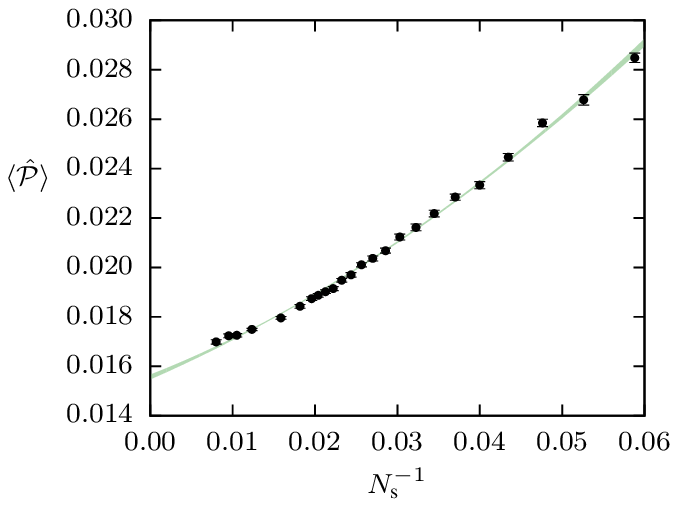}
\hfill\hfill\phantom.
\caption{\label{fig:volumeScaling010}Extrapolation of
$\langle\hat{\mathcal{P}}\rangle$ to infinite volume at fixed $\hat\lambda$ and
couplings $\hat f=100$ (left panel) and $\hat f=10$ (right panel). Depending on
the lattice size from $10^4$ up to $5\cdot 10^5$ configurations have been used.}
\end{figure}

Now that the finite temperature effects are under control the infinite volume
limit can be performed. Here, the infinite volume limit is taken
\emph{before} the continuum limit to finally work out the effect of the finite
lattice spacing. To accomplish an extrapolation the data at fixed finite lattice
spacing are fitted with functions of the form
\begin{equation}
\label{eq:n1:infiniteVolumeExtr}
\langle\hat{\mathcal{P}}\rangle(N_\text{s}) = A+
B N_\text{s}^{-1}+ C N_\text{s}^{-2}
\end{equation}
and extrapolated to $\Ns\to \infty$. For most of the couplings $\hat f$ 
lattices with $\Ns\in\{25,27,31,35,43,63\}$ are used. 
The two examples with
$\hat\lambda=0.1$ and $\hat f\in\{10,100\}$ in Fig.~\ref{fig:volumeScaling010}
illustrate the validity of the chosen extrapolation formula.

\begin{figure}\hfill
\includegraphics{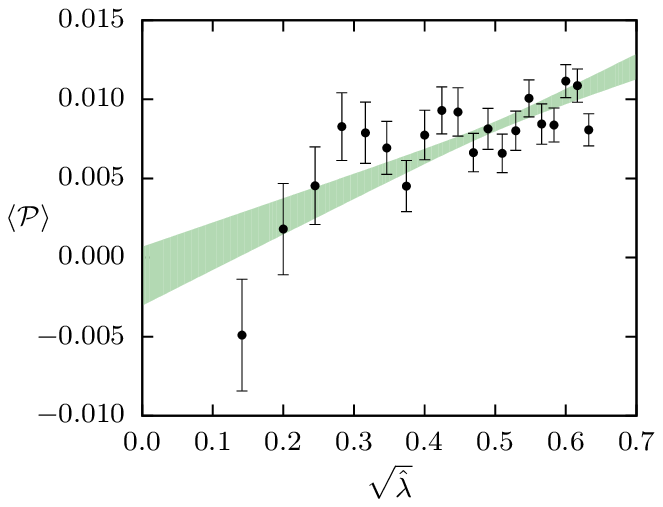}
\hfill
\includegraphics{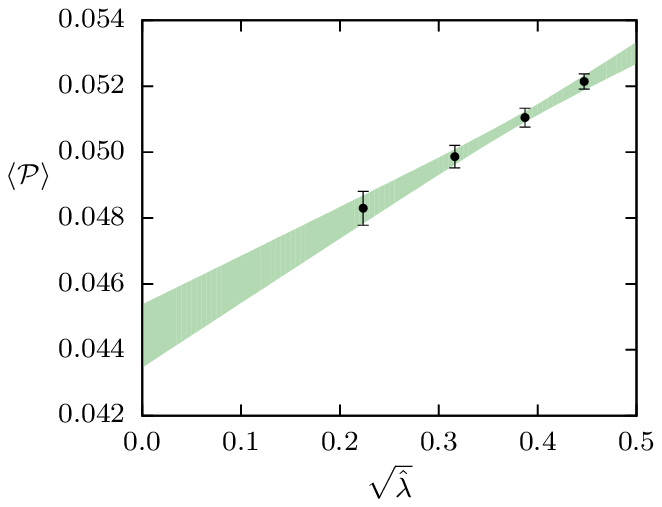}
\hfill\hfill\phantom.
\caption{\label{fig:continuumScaling10010}For the $\bbZ_2$ broken case
($f=100$, left panel) the Ward identity is fulfilled with
$\vev{\mathcal{P}}=-0.0012(19)$. For restored $\bbZ_2$ symmetry 
at $f=10$ (right panel) $\vev{\mathcal{P}}=0.0444(10)$ is obtained.}
\end{figure}

\begin{figure}
\centering
\includegraphics{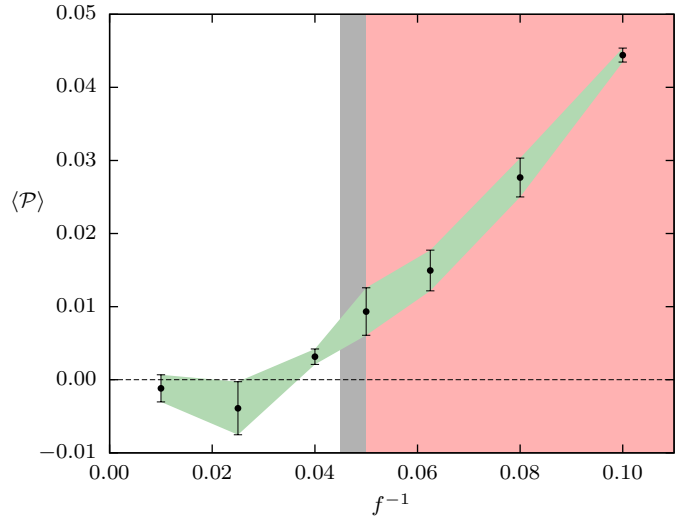}
\caption{\label{fig:prepot} Dimensionless prepotential $\mathcal{P}$ over
inverse coupling $f^{-1}$. The region shaded in red indicates the $\bbZ_2$
restored coupling range where supersymmetry is broken while the region shaded
in gray gives the error bound for the $\bbZ_2$ phase transition.}
\end{figure}

The last limit to be taken is the continuum limit. In the
simplest case corrections are of $\ord(a)$ and a linear extrapolation to the continuum limit is
possible. The extrapolation is done at each coupling
$f\in\{10,12.5,16,20,25,40,100\}$ and is shown for the extreme cases of
the present study, at $f=10$ and $f=100$, in Fig.~\ref{fig:continuumScaling10010}
where the validity of a linear extrapolation is visible.
In these cases a complete coincidence between
restored $\bbZ_2$ symmetry and spontaneously broken supersymmetry is visible.
All continuum extrapolated Ward identities in the considered coupling range
are plotted in Fig.~\ref{fig:prepot} and listed in
Tab.~\ref{tab:n1:prepot}.

The calculation of the dimensionless prepotential that serves as Ward identity
by Eq.~\eqref{eq:n1:prepot} has shown that after taking all necessary limiting
procedures supersymmetry is broken whenever the $\bbZ_2$ symmetry is
restored. Nevertheless, one inconsistency shows up for
$f$ slightly above the critical coupling for $\bbZ_2$
breaking (approximately $f\in[f_\text{c},27]$). In that region the $\bbZ_2$
symmetry is broken while the Ward identity is not fulfilled. This is in
contradiction with the fact that a broken $\bbZ_2$ symmetry strictly implies a
restored supersymmetry after all limits have been taken. Thus an analysis of
possible systematic errors is in order.

Firstly the used extrapolation formulae for the specific limits may not be
sufficient in every case. Secondly close to the critical
coupling at fixed $\hat\lambda$ where a second order phase transition (related
to an infinite continuum $\lambda$) with diverging correlation length occurs,  
the considered lattice volumes may still be too small to be in the applicability range for an
infinite volume extrapolation with Eq.~\eqref{eq:n1:infiniteVolumeExtr}. Thirdly
the ordering of limits may be of importance. As it has been found in the
supersymmetric quantum mechanics and the $\mathcal{N}=2$ Wess-Zumino model the
sign problem that arises from the fermionic part of the action becomes worse
for larger volumes at fixed lattice spacing and is weakened in the continuum
limit at fixed physical volume. Although the sign problem is completely absent
for small lattices $\sign\Pf M$ has not been
computed for the larger lattices due to the
numerical complexity of $\ord(N^3)$. Therefore sign problems cannot be excluded
for large lattice volumes and an extrapolation to infinite volume prior to the
continuum limit may turn out to be insufficient if no
reweighting with $\sign\Pf M$ is done.

\begin{table}
\centering
\begin{tabular}{c@{\hspace{5ex}}c}  \hline
$f$ & $\vev{\mathcal{P}}$ \\ \hline
$100$ & $-0.0012(19)$ \\
$40$  & $-0.0039(36)$ \\
$25$  & $0.0032(11)$ \\
$20$  & $0.0093(33)$ \\
$16$  & $0.0150(28)$ \\
$12.5$& $0.0277(27)$ \\
$10$  & $0.0444(10)$ \\ \hline
\end{tabular}
\caption{\label{tab:n1:prepot}Dimensionless
prepotential $\mathcal{P}$ after extrapolation to the limit of vanishing
temperature, infinite volume, and vanishing lattice spacing (in given order).}
\end{table}
Even with taking these possible systematic errors into account 
a supersymmetry breaking phase transition is confirmed where the corresponding
critical coupling coincides with that of the $\bbZ_2$ phase transition.
Nevertheless, the Binder cumulant technique for
the $\bbZ_2$ symmetry breaking provides a more reliable way to determine the
critical coupling because the extrapolation does not directly involve 
questionable extrapolation formulae.


\subsubsection{Masses}
\noindent
Ward identities are indicators for the restoration of
supersymmetry. Now, after the phase structure of the
theory is settled, further physical observables are of interest. 
Amongst them are the particle masses or the energy
difference between the ground state and the first excited state.
Here, one expects a fundamentally different behaviour of the masses in the
different phases \cite{Synatschke:2009nm}. In the supersymmetric phase a
degeneracy between the (finite) bosonic and fermionic mass is expected, similar to the $\mathcal{N}=2$ Wess-Zumino
model. For broken supersymmetry a \emph{goldstino} should arise as massless fermionic mode while the physical spacetime
volume serves as a regulator for the bosonic mass, which itself eventually vanishes in the infinite volume limit.

The analysis starts with the $\bbZ_2$ broken phase at a fixed coupling of
$f=100$. In this phase it is necessary to project the (finite volume) lattice
simulations onto one ground state to mimic the
suppression of tunnelling events in the infinite volume limit.\footnote{Here the configurations are projected
without loss of generality to the bosonic ground state.} This is the same technique as 
used in Sec.~\ref{chap:qm} for the SQM as well as for the $\mathcal{N}=2$ Wess-Zumino
model in \cite{Kastner:2008zc}, and it is necessary to finally
extrapolate the obtained masses to the infinite volume limit. Since a projection to one ground state is performed it
is \emph{not required} to stick to thermal boundary conditions as discussed in
Sec.~\ref{ssec:n1:z2breaking}. Thus, in order to remove the supersymmetry
breaking introduced by a finite temperature, periodic boundary conditions are
used also for the fermionic field to obtain correlators and masses in the
$\bbZ_2$ broken phase. Furthermore only square lattices are investigated to simplify the comparison of 
different lattice spacings and physical volumes.

\begin{figure}\hfill
\includegraphics{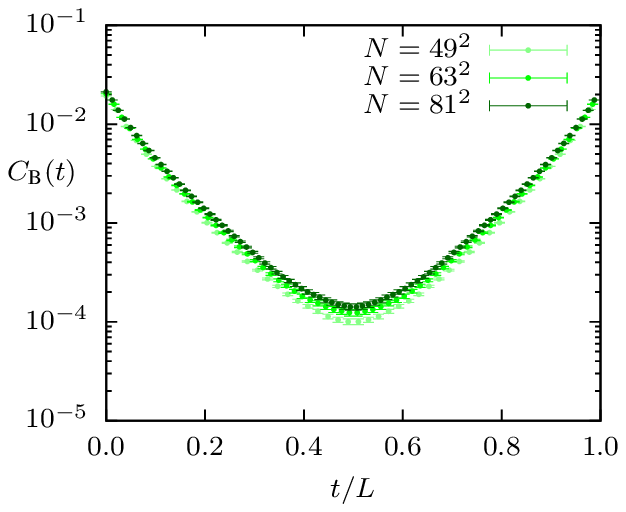}
\hfill
\includegraphics{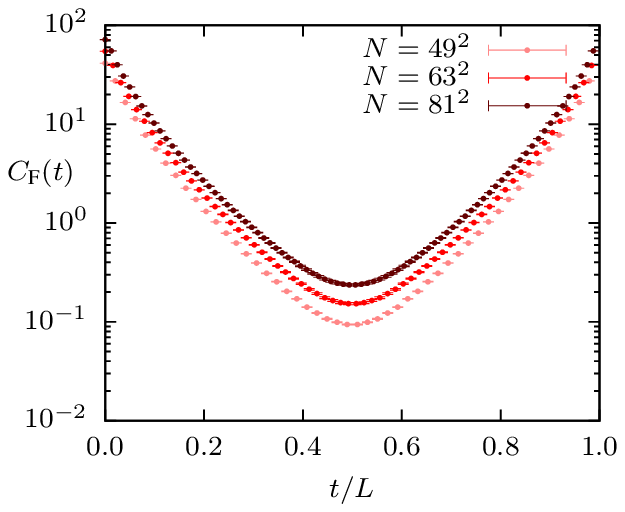}
\hfill\hfill\phantom.
\caption{\label{fig:n1:correlators100}Connected bosonic (left panel) and
fermionic (right panel) correlator in the $\bbZ_2$ broken phase ($f=100$) at fixed physical volume
$L\sqrt{\lambda} = 39.8$ for
different lattice spacings.}
\end{figure}%
\begin{table}
\begin{tabular*}{\textwidth}{c@{\extracolsep\fill}ccccc}  \hline
 & & $\Ns=49$ & $\Ns=63$ & $\Ns=81$ & cont.
  \\ \hline
$L\sqrt{\lambda}=19.9$ & bos.
 &$0.379(2)$   & $0.379(4)$  & $0.383(4)$  & $0.389(10)$\\
$L\sqrt{\lambda}=28.2$& bos.
 &$0.295(4)$  & $0.293(5)$  & $0.293(5)$  & $0.289(12)$\\
$L\sqrt{\lambda}=34.5$& bos.
 & $0.267(6)$ & $0.262(8)$  & $0.261(7)$  & $0.251(20)$\\
$L\sqrt{\lambda}=39.8$& bos.
 &$0.238(10)$ & $0.228(12)$ & $0.231(10)$ & $0.218(30)$\\
$L\sqrt{\lambda}=19.9$& ferm.
  &$0.300(1)$ & $0.294(2)$ & $0.291(2)$ & $0.277(4)$\\
$L\sqrt{\lambda}=28.2$& ferm.
 &$0.276(1)$ & $0.263(2)$ & $0.262(2)$ & $0.237(5)$\\
$L\sqrt{\lambda}=34.5$& ferm.
  &$0.270(1)$ & $0.261(2)$ & $0.252(2)$ & $0.225(5)$\\
$L\sqrt{\lambda}=39.8$& ferm.
  &$0.270(1)$ & $0.254(3)$ & $0.245(2)$ & $0.204(6)$\\ \hline
\end{tabular*}
\caption{\label{tab:n1:unbrokenMasses}Bosonic (upper rows) and fermionic (lower
rows) masses $m_{\text{B}/\text{F}}/\sqrt{\lambda}$ for fixed physical
volumes $L\sqrt{\lambda}$ and varying lattice spacing together with the
continuum extrapolation at coupling $f=100$.}
\end{table}
\begin{figure}\hfill
\includegraphics{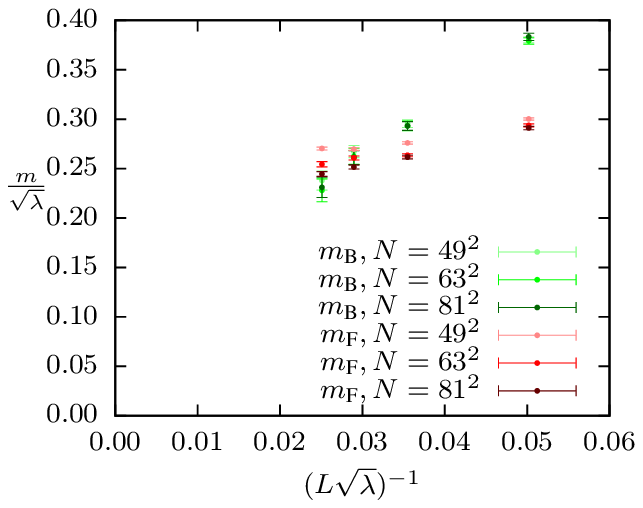}
\hfill
\includegraphics{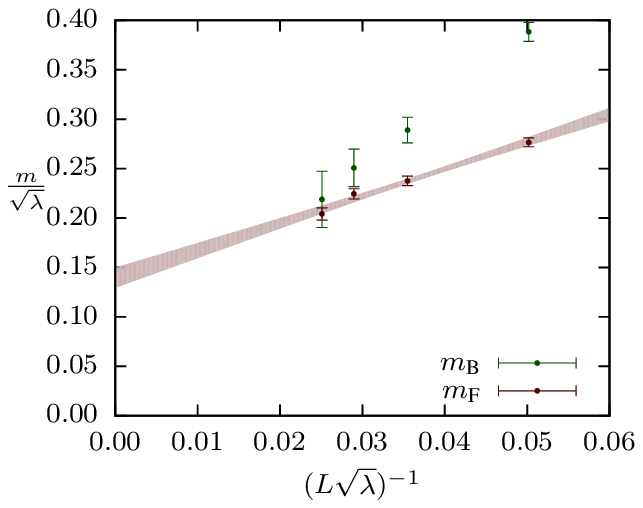}
\hfill\hfill\phantom.
\caption{\label{fig:n1:mass100}Left panel: Masses at finite lattice spacing at
coupling $f=100$. Right panel: Continuum extrapolated masses together with
the error bounds of an extrapolation of the fermionic mass to the infinite
volume limit (shaded area). $10^6$ configurations have been used for each data
point.}
\end{figure}%
Masses are extracted from
the correlators
\begin{equation}
C_\text{B}(t) = \Ns^{-2} \sum_{x,x'} \vev{\phi_{(t,x)} \phi_{(0,x')}} \quad
\text{and}\quad
C_\text{F}(t) = \Ns^{-2} \sum_{\alpha,x,x'} \vev{\bar\psi_{\alpha,(t,x)}
\psi_{\alpha,(0,x')}}
\end{equation}
using a $\cosh$ fit in a range $t\in
[L/3,2L/3]$. It is obvious that the
correlators at fixed physical volume depend on the lattice spacing (in Fig.~\ref{fig:n1:correlators100} the
fermionic correlator shows larger discretisation errors) and extracted masses
must therefore be extrapolated to the continuum limit. The continuum value is
reached via a linear extrapolation  that has already been used successfully
for the continuum extrapolation of results based on the SLAC derivative in unbroken supersymmetric quantum mechanics in \cite{Bergner:2007pu}. The results obtained
on lattices with $\Ns\in\{49,63,81\}$ for the bosonic and
fermionic masses at four different physical volumes, together with the continuum extrapolation, are given in Tab.~\ref{tab:n1:unbrokenMasses}
and are plotted in Fig.~\ref{fig:n1:mass100} (left panel) for finite lattice spacings.
Finally an infinite volume extrapolation of the continuum results is necessary.
Although boson masses approach the fermion masses at larger volumes, 
as predicted by supersymmetry, the statistical accuracy is not
sufficient for a reliable extrapolation. Therefore only the fermionic masses are
extrapolated linearly (see Fig.~\ref{fig:n1:mass100}, right panel) to the infinite volume
limit, resulting in $m_\text{F}/\sqrt{\lambda}=0.14(1)$. Unfortunately, the
bosonic mass can only be assumed to take the same limit. At this point 
a note concerning the finiteness of the bosonic mass is in order. The breaking 
of the \emph{discrete} $\bbZ_2$ symmetry does not lead to Goldstone
bosons and a finite mass is \emph{not excluded}. On the other hand, in the
$\bbZ_2$ symmetric phase the \emph{continuous supersymmetry} 
will be broken and massless goldstinos should appear in the spectrum of the theory.

\begin{figure}\hfill
\includegraphics{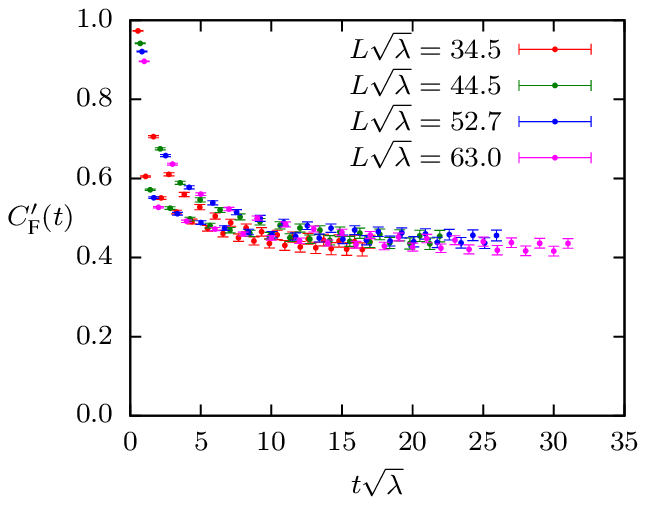}
\hfill
\includegraphics{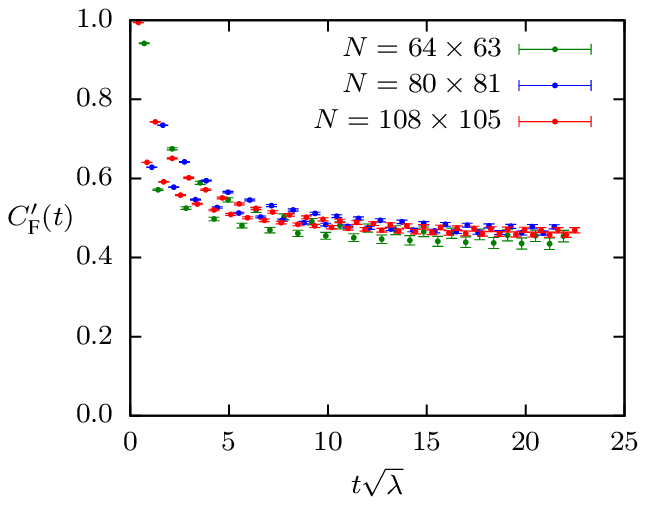}
\hfill\hfill\phantom.
\caption{\label{fig:n1:brokenCorrelatorF}Fermionic correlators at coupling $f=10$
in the thermal ensemble for different physical volumes on a $64\times 63$
lattice (left panel) and for different lattice sizes on a fixed physical volume $L\sqrt{\lambda}=44.5$ (right
panel). For symmetry reasons only the $t/L<0.5$ range is shown.}
\end{figure}%
The analysis of masses is therefore continued in the phase with broken
supersymmetry and restored $\bbZ_2$ symmetry. In this phase it is
mandatory to use thermal boundary conditions since bosonic and fermionic ground
state participate equally well in the path integral with unsuppressed tunnelling
even at infinite volume. Goldstinos will, similar to broken supersymmetric
quantum mechanics, show up as massless modes in
\begin{equation}
C'_\text{F}(t)=\Ns^{-2}
\sum_{\alpha,x,x'} \vev{\bar\psi_{1,(t,x)} \psi_{1,(0,x')}-\bar\psi_{2,(t,x)}
\psi_{2,(0,x')}},
\end{equation}
the spinor component combination that yields a $\cosh$ form
for thermal boundary conditions. At $\hat f=10$ the correlator has been computed
at varying physical volume with fixed lattice size and for fixed physical volume
with varying lattice spacing (see
Fig.~\ref{fig:n1:brokenCorrelatorF}).\footnote{The non-statistical fluctuations
showing up in the correlator can be traced back to the non-locality of the SLAC derivative. They
will decrease in the continuum limit, as visible in
Fig.~\ref{fig:n1:brokenCorrelatorF} (right panel).} A constant part of the
correlator is clearly visible and independent of the physical volume or lattice
spacing. This is an unambiguous sign of a goldstino.

\begin{figure}\hfill
\includegraphics{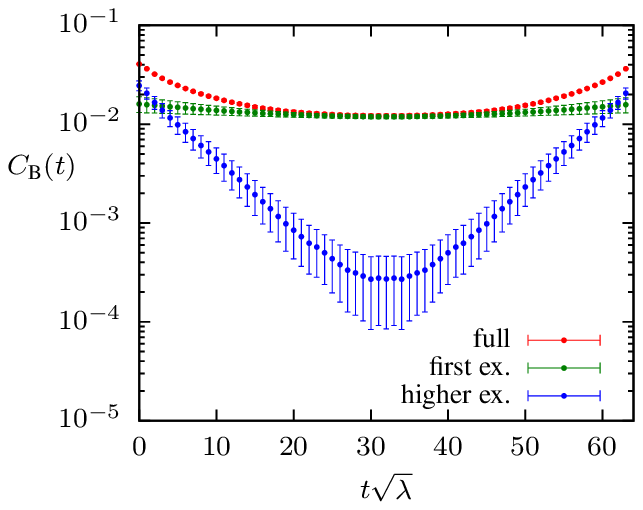}
\hfill
\includegraphics{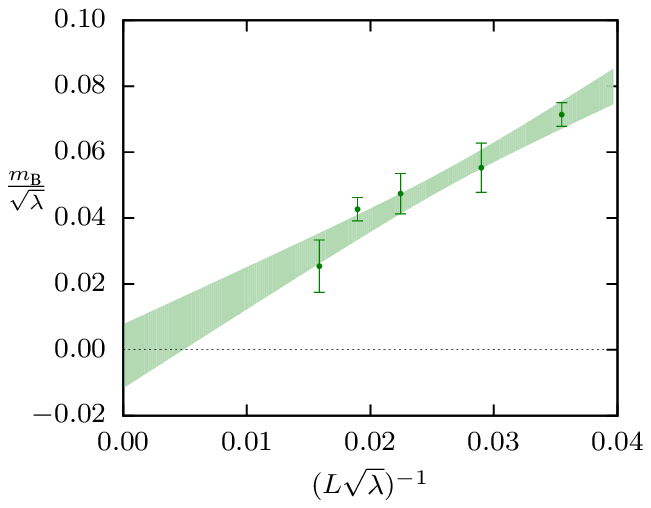}
\hfill\hfill\phantom.
\caption{\label{fig:n1:brokenCorrelatorB}Left panel: Bosonic correlator at
couping $f=10$ on a $64\times 63$ lattice for fixed physical volume
$L\sqrt{\lambda}=63$ where the contributions of first and higher excited states
have been separated. Right panel: Bosonic mass of the first excited state on
$64\times 63$ lattices and the infinite volume extrapolation (shaded area) for
$f=10$.}
\end{figure}

To get a complete picture the bosonic masses are calculated as well. These are
expected to vanish in the infinite volume limit as predicted by renormalisation
group methods in \cite{Synatschke:2009nm}. Again, the \emph{connected} bosonic
correlator at $f=10$ is computed and it is found to be composed of a part with nearly vanishing
mass, corresponding to the first excited state, and a part that arises from
higher excited states (see Fig.~\ref{fig:n1:brokenCorrelatorB}, left
panel).\footnote{The exponential decay of higher excited states is also visible
for $t\sqrt{\lambda}<10$ in the fermionic correlator, cf.\
Fig.~\ref{fig:n1:brokenCorrelatorF}.}  The masses of the first excited state are
now extrapolated (linearly) to infinite volume (see
Fig.~\ref{fig:n1:brokenCorrelatorB}, right panel). Here, only constant lattice sizes $N=64\times 63$
are used and the continuum limit is not performed. However, it has been checked for $L\sqrt{\lambda}=44.5$ on lattice sizes
$N=80\times 81$ and $N=108\times 105$ that the discretisation errors are still
below the statistical errors. The infinite volume extrapolation is in agreement
with $m_\text{B}\propto L^{-1}$ with an
extrapolated value of $m_\text{B}/\sqrt{\lambda}=-0.002(10)$, i.e.\ the bosonic mass
vanishes after the infrared regulator is removed, in agreement with
the results obtained by a functional renormalisation group approach
\cite{Synatschke:2009nm}.


\section{Conclusions}
\label{sec:conclusions}
\noindent
For the case of a supersymmetric quantum mechanics with dynamically broken
supersymmetry observables that are computed using a lattice regularisation with
the SLAC derivative completely coincide with results obtained from the
diagonalised Hamiltonian. The (bosonic/fermionic) nature of both ground states
can be explained with the impact of a change in boundary conditions on the
fermionic determinant. Correlators computed in the thermal ensemble show a
constant part for large $t$, which is a remnant of the degenerate ground states.
With a projection to one ground state the constant part is still visible in the
fermionic correlator, which goes at hand with the massless fermionic excitation
implied by the degeneracy. In the bosonic correlator the constant part vanishes
and the remaining exponential fall-off corresponds to the first excited state in
the bosonic spectrum.

On the level of Ward identities it is checked that the
ground state is not invariant under the supersymmetry and a simple Ward
identity is not fulfilled in the limit of vanishing lattice spacing and
temperature, as predicted from diagonalising the Hamiltonian.
In contrast to purely bosonic scalar models, the constraint effective potential
does not flatten out towards the conventional effective potential, which is
explained by the fact that no interpolating states between bosonic and fermionic
sector are accessibly in the finite temperature path integral.

We showed that it is possible to analyse supersymmetric quantum mechanics 
with broken supersymmetry with lattice methods based on the SLAC derivative.
The physical properties can be determined reliably and accurately. 
Supersymmetric quantum mechanics has been very useful to test 
sophisticated techniques that find applications in higher dimensional models.

The analysis of the $\mathcal{N}=(1,1)$ Wess-Zumino model in two dimensions
aimed at observing and understanding \emph{dynamical} supersymmetry breaking
from first principles. A lattice regularisation based on the SLAC derivative is used and
the choice of this regularisation is justified utilising the quenched model
where a complete agreement of the obtained critical coupling with the
reference value \cite{Schaich:2009jk} is found.

With this discretisation the $\bbZ_2$ symmetry breaking is analysed and a
renormalised continuum coupling is defined. For the first time a regulator
independent critical coupling is determined from lattice simulations. From the
computation of a Ward identity a complete coincidence between the restoration
of $\bbZ_2$ symmetry and the dynamical breaking of supersymmetry is obtained.
The computation of masses in the continuum limit for different physical volumes
completes the analysis and agrees with the picture of a finite and equal
bosonic and fermionic mass in the supersymmetric phase and the occurrence of a
massless goldstino for broken supersymmetry.

In future works it may be checked on the Ward identities by taking the infinite
volume limit \emph{after} the continuum limit has been carried out to suppress
possible systematic errors arising from the sign problem. Masses of higher excited states could be
within reach by using improved correlators. Finally a completely independent
calculation with different discretisation is
desirable. A formulation with Wilson fermions \cite{Golterman:1988ta}
is a natural choice, in particular because efficient lattice methods are
already available \cite{Baumgartner:2011cm}. However, in that case one must
ensure that a spontaneous breaking of the $\bbZ_2$ symmetry is not influenced
by the unavoidable explicit $\bbZ_2$ symmetry breaking induced by the Wilson 
mass term.

In both models the SLAC derivative has proven to be
successfully applicable, which can be traced back to the absence of gauge fields
in these models. Therefore it may be advantageous to consider the SLAC derivative as an
interesting alternative to Ginsparg-Wilson fermions also in simulations of the
four dimensional $\mathcal{N}=1$ Wess-Zumino model \cite{Chen:2010uc}.

Supersymmetrically improved lattice actions inevitably
include periodic boundary conditions for fermionic fields which lead to a severe
sign problem in supersymmetric theories with a dynamically
broken supersymmetry. Therefore the applicability of the improvement programme,
where one keeps part of the supersymmetry intact, becomes
questionable for these theories. It will be necessary to analyse if 
improved actions exist that give rise to a correct continuum limit without 
fine tuning even for thermal boundary conditions.


\section*{Acknowledgements}
\noindent
Numerous discussions with Tobias K\"astner, Georg
Bergner, and Franziska Synatschke are gratefully acknowledged. CW thanks for the
support by the Studienstiftung des deutschen Volkes. This work has been
supported by the DFG Research Training Group ``Quantum and Gravitational
Fields'' GRK~1523 and the DFG grant Wi~777/10-1. 
The simulations have been carried out at the Omega cluster of the TPI.

\bibliographystyle{hepbib}
\bibliography{literature}

\end{document}